\documentclass[12pt]{article}
\usepackage{amsmath,amssymb,amsthm,amsxtra,overpic,bbm,bm,epsfig,ulem}
\usepackage{color}
\usepackage{cite}

\usepackage{graphicx}

\usepackage{hyperref}
\usepackage{url}
\usepackage{multirow}
\usepackage{textcomp}

\def\thefootnote{\fnsymbol{footnote}}

\begin{document}

\vspace{0.2cm}

\begin{center}
{\large\bf A full parametrization of the $9\times 9$ active-sterile flavor mixing matrix in
the inverse or linear seesaw scenario of massive neutrinos}
\end{center}

\vspace{0.2cm}

\begin{center}
{\bf He-chong Han$^{1,2}$}
\footnote{E-mail: hanhechong@ihep.ac.cn},
{\bf Zhi-zhong Xing$^{1,2,3}$}
\footnote{E-mail: xingzz@ihep.ac.cn}
\\
{\small
$^{1}$Institute of High Energy Physics, Chinese Academy of Sciences, Beijing 100049, China \\
$^{2}$School of Physical Sciences,
University of Chinese Academy of Sciences, Beijing 100049, China \\
$^{3}$Center of High Energy Physics, Peking University, Beijing 100871, China}
\end{center}

\vspace{2cm}
\begin{abstract}
The inverse and linear seesaw scenarios are two typical extensions of the canonical
seesaw mechanism, which contain
much more sterile degrees of freedom but can naturally
explain the smallness of three active neutrino masses at a sufficiently low energy
scale (e.g., the TeV scale). To fully describe the mixing among three active neutrinos,
three sterile neutrinos and three extra gauge-singlet neutral fermions in either of
these two seesaw paradigms, we present the {\it first} full parametrization
of the $9\times 9$ flavor mixing matrix in terms of 36 rotation angles
and 36 CP-violating phases. The exact inverse and linear seesaw formulas are
derived, respectively; and possible deviations of the $3\times 3$ active neutrino
mixing matrix from its unitary limit are discussed by calculating the effective
Jarlskog invariants and unitarity nonagons.
\end{abstract}

\newpage

\def\thefootnote{\arabic{footnote}}
\setcounter{footnote}{0}
\setcounter{figure}{0}

\section{Introduction}

The absence of the right-handed neutrino fields in the standard model (SM) makes
it impossible to accommodate a gauge-invariant neutrino mass term like the
Yukawa-interaction term for the charged leptons \cite{Weinberg:1967tq}. Therefore,
the simplest way to go beyond the SM and generate finite masses for three
active neutrinos (i.e., $\nu^{}_e$, $\nu^{}_\mu$ and $\nu^{}_\tau$) is to
introduce three right-handed neutrino fields and allow for the gauge-invariant
{\it neutrino} Yukawa interactions. But such a Dirac neutrino mass term has no
way to explain why the active neutrinos are so light as compared with the respective
charged leptons, which participate in the standard weak interactions in the
same way. Moreover, the right-handed neutrino fields and their charge-conjugate
counterparts (which are left-handed) can form a gauge-invariant but
lepton-number-violating Majorana mass term, which should not be discarded
in general. Combining this Majorana neutrino
mass term with the aforementioned Dirac mass term
leads us to the canonical (type-I) seesaw mechanism \cite{Minkowski:1977sc,
Yanagida:1979as,GellMann:1980vs,Glashow:1979nm,Mohapatra:1979ia}, in which
{\it all the six} neutrino mass eigenstates have the Majorana nature (i.e., their
charge conjugates are equal to themselves; or equivalently, the neutrinos
are their own antiparticles \cite{Majorana:1937vz}) and the smallness of three
active neutrino masses $m^{}_i$ can naturally be attributed to the largeness of
three sterile neutrino masses $M^{}_i$ (for $i = 1,2,3$). Note that $M^{}_i$
act as the cut-off scales in the SM effective field theory with a unique
dimension-five Weinberg operator for generating tiny masses of the active
neutrinos \cite{Weinberg:1979sa}, and the ``fulcrum" of such a seesaw is
just around the electroweak scale $\Lambda^{}_{\rm EW} \sim 10^2$ GeV.
Unfortunately, the naturalness prerequisite $M^{}_i \gg \Lambda^{}_{\rm EW}$
implies that this conventional seesaw picture has essentially lost its
testability in any feasible high-energy experiments \cite{Xing:2009in}.

In this regard a possible way out is to lower the seesaw scale (i.e., the values
of $M^{}_i$) by introducing more extra degrees of freedom, and the typical
examples of this kind include the inverse (or double) seesaw scenario
\cite{Mohapatra:1986bd,Wyler:1982dd} and the linear seesaw scenario
\cite{Wyler:1982dd,Akhmedov:1995ip,Akhmedov:1995vm,Barr:2003nn}
\footnote{A natural extension of the inverse seesaw scenario to the multiple
seesaw scenario has been proposed in Ref.~\cite{Xing:2009hx}.}.
Besides the left-handed neutrino fields $\nu^{}_{\alpha {\rm L}}$ and the
right-handed neutrino fields $N^{}_{\alpha {\rm R}}$ (for $\alpha = e, \mu, \tau$),
the inverse or linear seesaw mechanism requires the adding of three neutral SM
gauge-singlet fermions $S^{}_{\alpha {\rm R}}$ (for $\alpha = e, \mu, \tau$)
and one scalar singlet $\Phi$. With such new degrees of freedom, a generic
gauge-invariant neutrino mass term can be written as follows:
\begin{eqnarray}
-{\cal L}^{}_{\rm mass} = \overline{\ell^{}_{\rm L}} Y^{}_\nu \widetilde{H}
N^{}_{\rm R} + \frac{1}{2} \overline{(N^{}_{\rm R})^c} M^{}_{\rm R} N^{}_{\rm R}
+ \overline{\nu^{}_{\rm L}} Y^{\prime}_\nu \Phi S^{}_{\rm R} +
\overline{(N^{}_{\rm R})^c} Y^{}_S \Phi S^{}_{\rm R} +
\frac{1}{2} \overline{(S^{}_{\rm R})^c} \mu S^{}_{\rm R} + {\rm h.c.} \; ,
\end{eqnarray}
where
$\nu^{}_{\rm L} = (\nu^{}_{e \rm L} \hspace{0.15cm} \nu^{}_{\mu \rm L}
\hspace{0.15cm} \nu^{}_{\tau \rm L})^T$,
$N^{}_{\rm R} = (N^{}_{e \rm R} \hspace{0.15cm} N^{}_{\mu \rm R}
\hspace{0.15cm} N^{}_{\tau \rm R})^T$ and
$S^{}_{\rm R} = (\nu^{}_{e \rm R} \hspace{0.15cm} \nu^{}_{\mu \rm R}
\hspace{0.15cm} \nu^{}_{\tau \rm R})^T$
stand respectively for the vector columns of left-handed neutrino fields,
right-handed neutrino fields and gauge-singlet fermion fields,
$\ell^{}_{\rm L}$ denotes the lepton doublet of the SM, $\widetilde{H}$ is
defined as $\widetilde{H} \equiv {\rm i} \sigma^{}_2 H^*$ with $H$ being the
Higgs doublet of the SM, and the mass matrices $M^{}_{\rm R}$ and $\mu$ are
both symmetric. After spontaneous symmetry breaking, Eq.~(1) turns out to be
\begin{eqnarray}
-{\cal L}^{\prime}_{\rm mass} = \frac{1}{2}
\overline{\left[\nu^{}_{\rm L} \hspace{0.2cm} (N^{}_{\rm R})^c
\hspace{0.2cm} (S^{}_{\rm R})^c\right]}
\left( \begin{matrix} 0 & M^{}_{\rm D} & \varepsilon
\cr M^T_{\rm D} & M^{}_{\rm R} & M^{}_S \cr \varepsilon^{T} & M^T_S &
\mu \cr \end{matrix} \right) \left( \begin{matrix} (\nu^{}_{\rm L})^c \cr
N^{}_{\rm R} \cr S^{}_{\rm R} \cr \end{matrix} \right) + {\rm h.c.} \; ,
\end{eqnarray}
where $M^{}_{\rm D} = Y^{}_\nu \langle H\rangle$,
$\varepsilon = Y^\prime_\nu \langle \Phi\rangle$ and
$M^{}_S = Y^{}_S \langle \Phi\rangle$. Switching off the extra neutral
gauge-singlet fermion fields, we are immediately left with the neutrino
mass term of the canonical seesaw mechanism from Eqs.~(1) and (2).
To partly reduce the number of new free parameters associated with the overall
$9\times 9$ neutrino mass matrix in Eq.~(2), one has considered the
following two simplified but phenomenologically interesting cases.
\begin{itemize}
\item     The {\it inverse} seesaw scenario with $M^{}_{\rm R} = 0$ and
$\varepsilon = 0$ \cite{Mohapatra:1986bd,Wyler:1982dd}, where the
mass scale of $\mu$ is naturally small because it is the
only lepton-number-violating term, and the mass scale
of $M^{}_S$ can be considerably larger than that of $M^{}_{\rm D}$
if $\langle \Phi\rangle \gg \langle H\rangle$ and $Y^{}_S \sim Y^{}_\nu$
are taken. The effective Majorana neutrino mass matrix
for three active neutrinos is therefore given by
\begin{eqnarray}
M^{}_\nu \simeq M^{}_{\rm D} (M^T_S)^{-1} \mu \hspace{0.06cm}
(M^{}_S)^{-1} M^T_{\rm D} \;
\end{eqnarray}
in the leading-order approximation. So the tiny mass eigenvalues of
$M^{}_\nu$ are mainly attributed to the smallness of $\mu$, and they are
further suppressed by the largeness of $M^{}_S$ as compared with
$M^{}_{\rm D}$ even if $M^{}_S \sim {\cal O}(1)$ TeV is assumed.

\item     The {\it linear} seesaw scenario with $M^{}_{\rm R} = 0$ and $\mu = 0$
\cite{Wyler:1982dd,Akhmedov:1995ip,Akhmedov:1995vm,Barr:2003nn}, where
$\varepsilon$ is the only lepton-number-violating term, and thus its mass
scale is naturally small. In this case the effective Majorana neutrino
mass matrix for three active neutrinos is found to be
\begin{eqnarray}
M^{}_\nu \simeq - \varepsilon M^{-1}_{S} M^T_{\rm D} -
(\varepsilon M^{-1}_{S} M^T_{\rm D})^T \;
\end{eqnarray}
in the leading-order approximation. As a result, the smallness of $M^{}_\nu$
is mainly ascribed to that of $\varepsilon$ and further suppressed by the
ratio of the mass scales of $M^{}_{\rm D}$ to $M^{}_S$.
\end{itemize}
In either case the seesaw scale can be successfully lowered to the TeV scale
which is experimentally accessible at the Large Hadron Collider (LHC).

Although a lot of work has been done to explore various phenomenological
consequences of the inverse or linear seesaw scenario (see, e.g.,
Refs.~\cite{Malinsky:2009df,Hirsch:2009mx,Bergstrom:2010qb,Dev:2009aw,
Hettmansperger:2011bt,Das:2012ze,
BhupalDev:2012jvh,Deppisch:2015cua,Sinha:2015ooa,CarcamoHernandez:2019iwh,
Camara:2020efq,Cao:2021lmj,Verma:2021koo,Zhang:2021olk,Hagedorn:2021ldq}),
a complete description of the $9\times 9$ active-sterile flavor mixing matrix for
such a seesaw picture has been lacking. Following the full Euler-like parametrization
of active-sterile flavor mixing in the type-I or type-(I+II) seesaw mechanism
with three sterile neutrinos done previously by one of us
\cite{Xing:2011ur,Xing:2007zj,Xing:2019vks},
here we are going to present the {\it first} full parametrization of the $9\times 9$
flavor mixing matrix in terms of 36 Euler rotation angles
and 36 CP-violating phases in the inverse or linear seesaw scenario.
The exact inverse or linear seesaw formula will also be
derived; and possible deviations of the $3\times 3$ active neutrino
mixing matrix (i.e., the so-called Pontecorvo-Maki-Nakagawa-Sakata lepton flavor
mixing matrix \cite{Pontecorvo:1957cp,Maki:1962mu,Pontecorvo:1967fh})
from its unitary limit will be discussed by calculating the effective
Jarlskog invariants and unitarity nonagons. This study is expected to be
useful for generally describing possible interplays between three active neutrino
species and up to six species of extra (light or heavy) degrees of freedom no
matter whether a specific seesaw scenario is taken into account or not.

The remaining parts of this paper are organized as follows. In section 2 we
parameterize the $9\times 9$ flavor mixing matrix with 36 Euler rotation angles
and 36 CP-violating phases in such a way that the primary unitary $3\times 3$
flavor mixing submatrices of three active neutrinos, three sterile neutrinos and
three extra neutral fermions are linked and modified by the intermediate flavor
mixing matrices. Section 3 is devoted to deriving the exact inverse or linear
seesaw formula and reproducing its approximate expression shown in
Eq.~(3) or Eq.~(4). In section 4 we examine possible deviations of the $3\times 3$
PMNS matrix from its unitary limit by calculating the effective
Jarlskog invariants and discussing the unitarity nonagons. A brief summary of our
main results, together with some further discussions, is finally made in section 5.

\section{Parametrization of the $9\times 9$ flavor mixing matrix}

To fully describe the parameter space of flavor mixing between three active neutrinos
and their six sterile counterparts, one may write out the $9\times 9$ active-sterile
flavor mixing matrix $\cal U$ in terms of 36 two-dimensional unitary matrices
$O^{}_{ij}$ ($1 \leq i < j \leq 9$) as follows:
\begin{eqnarray}
{\cal U} \hspace{-0.2cm} & = & \hspace{-0.2cm}
\big(O^{}_{89} O^{}_{79} O^{}_{69} O^{}_{59} O^{}_{49} O^{}_{39} O^{}_{29} O^{}_{19}\big)
\big(O^{}_{78} O^{}_{68} O^{}_{58} O^{}_{48} O^{}_{38} O^{}_{28} O^{}_{18}\big)
\big(O^{}_{67} O^{}_{57} O^{}_{47} O^{}_{37} O^{}_{27} O^{}_{17}\big)
\nonumber \\
\hspace{-0.2cm} & & \hspace{-0.2cm}
\big(O^{}_{56} O^{}_{46} O^{}_{36} O^{}_{26} O^{}_{16}\big)
\big(O^{}_{45} O^{}_{35} O^{}_{25} O^{}_{15}\big)
\big(O^{}_{34} O^{}_{24} O^{}_{14}\big)
\big(O^{}_{23} O^{}_{13} O^{}_{12}\big) \; ,
\end{eqnarray}
where only the combination $\big(O^{}_{23} O^{}_{13} O^{}_{12}\big)$ is purely
associated with the flavor mixing sector of three active neutrinos.
Following the example of Refs.~\cite{Xing:2011ur,Xing:2007zj,Xing:2019vks},
let us adjust the ordering of $O^{}_{ij}$ in $\cal U$ in such a way that
\begin{eqnarray}
{\cal U} \hspace{-0.2cm} & = & \hspace{-0.2cm}
\left( \begin{matrix} I & 0 & 0 \cr 0 & I & 0 \cr 0 & 0 & S^{}_0
\end{matrix} \right) L^{}_3
\left( \begin{matrix} I & 0 & 0 \cr 0 & U^{\prime}_0 & 0 \cr 0 & 0 & I
\end{matrix} \right) L^{}_2 L^{}_1
\left( \begin{matrix} U^{}_0 & 0 & 0 \cr 0 & I & 0 \cr 0 & 0 & I
\end{matrix} \right) \; ,
\end{eqnarray}
where $I$ denotes the $3\times 3$ identity matrix; the $3\times 3$ unitary matrices
\begin{eqnarray}
U^{}_0 = O^{}_{23} O^{}_{13} O^{}_{12} \; ,
\quad
U^\prime_0 = O^{}_{56} O^{}_{46} O^{}_{45} \; ,
\quad
S^{}_0 = O^{}_{89} O^{}_{79} O^{}_{78}
\end{eqnarray}
describe the primary flavor mixing sectors of three active neutrinos, three
sterile neutrinos and three extra neutral fermions, respectively; and
the $9\times 9$ matrices
\begin{eqnarray}
L^{}_1 \hspace{-0.2cm} & = & \hspace{-0.2cm}
O^{}_{36} O^{}_{26} O^{}_{16} O^{}_{35} O^{}_{25} O^{}_{15} O^{}_{34} O^{}_{24} O^{}_{14}
= \left( \begin{matrix} A^{}_1 & R^{}_1 & 0 \cr S^{}_1 & B^{}_1 & 0 \cr 0 & 0 & I
\end{matrix} \right) \; ,
\nonumber \\
L^{}_2 \hspace{-0.2cm} & = & \hspace{-0.2cm}
O^{}_{39} O^{}_{29} O^{}_{19} O^{}_{38} O^{}_{28} O^{}_{18} O^{}_{37} O^{}_{27} O^{}_{17}
= \left( \begin{matrix} A^{}_2 & 0 & R^{}_2 \cr 0 & I & 0 \cr S^{}_2 & 0 & B^{}_2
\end{matrix} \right) \; ,
\nonumber \\
L^{}_3 \hspace{-0.2cm} & = & \hspace{-0.2cm}
O^{}_{69} O^{}_{59} O^{}_{49} O^{}_{68} O^{}_{58} O^{}_{48} O^{}_{67} O^{}_{57} O^{}_{47}
= \left( \begin{matrix} I & 0 & 0 \cr 0 & A^{}_3 & R^{}_3 \cr 0 & S^{}_3 & B^{}_3
\end{matrix} \right) \; \hspace{0.7cm}
\end{eqnarray}
describe the interplays between any two of the three sectors. There are two obvious
advantages associated with the parametrization of $\cal U$ in Eq.~(6): on the one hand,
the flavor mixing angles of $L^{}_1$ and $L^{}_2$ are naturally small because they
measure the strength of interplay between the active sector and two sterile sectors;
on the other hand, the three flavor sectors will automatically become decoupled if
the off-diagonal elements of $L^{}_1$, $L^{}_2$ and $L^{}_3$ are all switched off.

To be more specific, the three primary $3\times 3$ unitary flavor mixing matrices
are given by
\begin{eqnarray}
U^{}_0 \hspace{-0.2cm} & = & \hspace{-0.2cm}
\left( \begin{matrix} c^{}_{12} c^{}_{13} & \hat{s}^*_{12}
	c^{}_{13} & \hat{s}^*_{13} \cr -\hat{s}^{}_{12} c^{}_{23} -
	c^{}_{12} \hat{s}^{}_{13} \hat{s}^*_{23} & c^{}_{12} c^{}_{23} -
	\hat{s}^*_{12} \hat{s}^{}_{13} \hat{s}^*_{23} & c^{}_{13}
	\hat{s}^*_{23} \cr \hat{s}^{}_{12} \hat{s}^{}_{23} - c^{}_{12}
	\hat{s}^{}_{13} c^{}_{23} & ~ -c^{}_{12} \hat{s}^{}_{23} -
	\hat{s}^*_{12} \hat{s}^{}_{13} c^{}_{23} ~ & c^{}_{13} c^{}_{23}
\end{matrix} \right) \;,
\nonumber \\
U^{\prime}_0 \hspace{-0.2cm} & = & \hspace{-0.2cm}
\left( \begin{matrix} c^{}_{45} c^{}_{46} & \hat{s}^*_{45}
	c^{}_{46} & \hat{s}^*_{46} \cr -\hat{s}^{}_{45} c^{}_{56} -
	c^{}_{45} \hat{s}^{}_{46} \hat{s}^*_{56} & c^{}_{45} c^{}_{56} -
	\hat{s}^*_{45} \hat{s}^{}_{46} \hat{s}^*_{56} & c^{}_{46}
	\hat{s}^*_{56} \cr \hat{s}^{}_{45} \hat{s}^{}_{56} - c^{}_{45}
	\hat{s}^{}_{46} c^{}_{56} & ~ -c^{}_{45} \hat{s}^{}_{56} -
	\hat{s}^*_{45} \hat{s}^{}_{46} c^{}_{56} ~ & c^{}_{46} c^{}_{56}
\end{matrix} \right) \;,
\nonumber \\
S^{}_0 \hspace{-0.2cm} & = & \hspace{-0.2cm}
\left( \begin{matrix} c^{}_{78} c^{}_{79} & \hat{s}^*_{78}
	c^{}_{79} & \hat{s}^*_{79} \cr -\hat{s}^{}_{78} c^{}_{89} -
	c^{}_{78} \hat{s}^{}_{79} \hat{s}^*_{89} & c^{}_{78} c^{}_{89} -
	\hat{s}^*_{78} \hat{s}^{}_{79} \hat{s}^*_{89} & c^{}_{79}
	\hat{s}^*_{89} \cr \hat{s}^{}_{78} \hat{s}^{}_{89} - c^{}_{78}
	\hat{s}^{}_{79} c^{}_{89} & ~ -c^{}_{78} \hat{s}^{}_{89} -
	\hat{s}^*_{78} \hat{s}^{}_{79} c^{}_{89} ~ & c^{}_{79} c^{}_{89}
\end{matrix} \right)\; ,
\end{eqnarray}
where $c^{}_{ij} \equiv \cos\theta^{}_{ij}$ and $\hat{s}^{}_{ij} \equiv
e^{{\rm i}\delta^{}_{ij}} \sin\theta^{}_{ij}$ with $\theta^{}_{ij}$ and
$\delta^{}_{ij}$ (for $1 \leq i < j \leq 9$)
standing respectively for the rotation angles and phase angles. Moreover,
the four $3\times 3$ submatrices of $L^{}_1$ read as
\begin{eqnarray}
A^{}_1 \hspace{-0.2cm} & = & \hspace{-0.2cm}
\left( \begin{matrix} c^{}_{14} c^{}_{15} c^{}_{16} & 0 & 0
	\cr\vspace{-0.4cm}\cr
	\begin{array}{l} -c^{}_{14} c^{}_{15} \hat{s}^{}_{16} \hat{s}^*_{26} -
	c^{}_{14} \hat{s}^{}_{15} \hat{s}^*_{25} c^{}_{26} \\
	-\hat{s}^{}_{14} \hat{s}^*_{24} c^{}_{25} c^{}_{26} \end{array} &	
c^{}_{24} c^{}_{25} c^{}_{26} & 0 \cr\vspace{-0.4cm}\cr
	\begin{array}{l} -c^{}_{14} c^{}_{15} \hat{s}^{}_{16} c^{}_{26} \hat{s}^*_{36}
	+ c^{}_{14} \hat{s}^{}_{15} \hat{s}^*_{25} \hat{s}^{}_{26} \hat{s}^*_{36} \\
	- c^{}_{14} \hat{s}^{}_{15} c^{}_{25} \hat{s}^*_{35} c^{}_{36} +
	\hat{s}^{}_{14} \hat{s}^*_{24} c^{}_{25} \hat{s}^{}_{26}
	\hat{s}^*_{36} \\
	+ \hat{s}^{}_{14} \hat{s}^*_{24} \hat{s}^{}_{25} \hat{s}^*_{35}
	c^{}_{36} - \hat{s}^{}_{14} c^{}_{24} \hat{s}^*_{34} c^{}_{35}
	c^{}_{36} \end{array} &
	\begin{array}{l} -c^{}_{24} c^{}_{25} \hat{s}^{}_{26} \hat{s}^*_{36} -
	c^{}_{24} \hat{s}^{}_{25} \hat{s}^*_{35} c^{}_{36} \\
	-\hat{s}^{}_{24} \hat{s}^*_{34} c^{}_{35} c^{}_{36} \end{array} &	
c^{}_{34} c^{}_{35} c^{}_{36} \cr \end{matrix} \right) \;,
\nonumber \\
B^{}_1 \hspace{-0.2cm} & = & \hspace{-0.2cm}
\left( \begin{matrix} c^{}_{14} c^{}_{24} c^{}_{34} & 0 & 0 \cr\vspace{-0.4cm}\cr
	\begin{array}{l} -c^{}_{14} c^{}_{24} \hat{s}^{*}_{34} \hat{s}^{}_{35} -
	c^{}_{14} \hat{s}^{*}_{24} \hat{s}^{}_{25} c^{}_{35} \\
	-\hat{s}^{*}_{14} \hat{s}^{}_{15} c^{}_{25} c^{}_{35} \end{array} &
	c^{}_{15} c^{}_{25} c^{}_{35} & 0 \cr\vspace{-0.4cm}\cr
	\begin{array}{l} -c^{}_{14} c^{}_{24} \hat{s}^{*}_{34} c^{}_{35} \hat{s}^{}_{36}
	+ c^{}_{14} \hat{s}^{*}_{24} \hat{s}^{}_{25} \hat{s}^{*}_{35} \hat{s}^{}_{36} \\
	- c^{}_{14} \hat{s}^{*}_{24} c^{}_{25} \hat{s}^{}_{26} c^{}_{36} +
	\hat{s}^{*}_{14} \hat{s}^{}_{15} c^{}_{25} \hat{s}^{*}_{35}
	\hat{s}^{}_{36} \\
	+ \hat{s}^{*}_{14} \hat{s}^{}_{15} \hat{s}^{*}_{25} \hat{s}^{}_{26}
	c^{}_{36} - \hat{s}^{*}_{14} c^{}_{15} \hat{s}^{}_{16} c^{}_{26}
	c^{}_{36} \end{array} &
	\begin{array}{l} -c^{}_{15} c^{}_{25} \hat{s}^{*}_{35} \hat{s}^{}_{36} -
	c^{}_{15} \hat{s}^{*}_{25} \hat{s}^{}_{26} c^{}_{36} \\
	-\hat{s}^{*}_{15} \hat{s}^{}_{16} c^{}_{26} c^{}_{36} \end{array} &
	c^{}_{16} c^{}_{26} c^{}_{36} \cr \end{matrix} \right) \;;
\hspace{0.9cm}
\end{eqnarray}
and
\begin{eqnarray}
R^{}_1 \hspace{-0.2cm} & = & \hspace{-0.2cm}
\left( \begin{matrix} \hat{s}^*_{14} c^{}_{15} c^{}_{16} &
	\hat{s}^*_{15} c^{}_{16} & \hat{s}^*_{16} \cr\vspace{-0.4cm}\cr
	\begin{array}{l} -\hat{s}^*_{14} c^{}_{15} \hat{s}^{}_{16} \hat{s}^*_{26} -
	\hat{s}^*_{14} \hat{s}^{}_{15} \hat{s}^*_{25} c^{}_{26} \\
	+ c^{}_{14} \hat{s}^*_{24} c^{}_{25} c^{}_{26} \end{array} & -
	\hat{s}^*_{15} \hat{s}^{}_{16} \hat{s}^*_{26} + c^{}_{15}
	\hat{s}^*_{25} c^{}_{26} & c^{}_{16} \hat{s}^*_{26} \cr\vspace{-0.4cm}\cr
	\begin{array}{l} -\hat{s}^*_{14} c^{}_{15} \hat{s}^{}_{16} c^{}_{26}
	\hat{s}^*_{36} + \hat{s}^*_{14} \hat{s}^{}_{15} \hat{s}^*_{25}
	\hat{s}^{}_{26} \hat{s}^*_{36} \\ - \hat{s}^*_{14} \hat{s}^{}_{15}
	c^{}_{25} \hat{s}^*_{35} c^{}_{36} - c^{}_{14} \hat{s}^*_{24}
	c^{}_{25} \hat{s}^{}_{26}
	\hat{s}^*_{36} \\
	- c^{}_{14} \hat{s}^*_{24} \hat{s}^{}_{25} \hat{s}^*_{35}
	c^{}_{36} + c^{}_{14} c^{}_{24} \hat{s}^*_{34} c^{}_{35} c^{}_{36}
	\end{array} &
	\begin{array}{l} -\hat{s}^*_{15} \hat{s}^{}_{16} c^{}_{26} \hat{s}^*_{36}
	- c^{}_{15} \hat{s}^*_{25} \hat{s}^{}_{26} \hat{s}^*_{36} \\
	+c^{}_{15} c^{}_{25} \hat{s}^*_{35} c^{}_{36} \end{array} &
	c^{}_{16} c^{}_{26} \hat{s}^*_{36} \cr \end{matrix} \right) \;,
\nonumber \\
S^{}_1 \hspace{-0.2cm} & = & \hspace{-0.2cm}
\left( \begin{matrix} -\hat{s}^{}_{14} c^{}_{24} c^{}_{34} &
	-\hat{s}^{}_{24} c^{}_{34} & -\hat{s}^{}_{34} \cr\vspace{-0.4cm}\cr
	\begin{array}{l} \hat{s}^{}_{14} c^{}_{24} \hat{s}^{*}_{34} \hat{s}^{}_{35}
	+ \hat{s}^{}_{14} \hat{s}^{*}_{24} \hat{s}^{}_{25} c^{}_{35} \\
	- c^{}_{14} \hat{s}^{}_{15} c^{}_{25} c^{}_{35} \end{array} &
	\hat{s}^{}_{24} \hat{s}^{*}_{34} \hat{s}^{}_{35} - c^{}_{24}
	\hat{s}^{}_{25} c^{}_{35} & -c^{}_{34} \hat{s}^{}_{35} \cr\vspace{-0.4cm}\cr
	\begin{array}{l} \hat{s}^{}_{14} c^{}_{24} \hat{s}^{*}_{34} c^{}_{35}
	\hat{s}^{}_{36} - \hat{s}^{}_{14} \hat{s}^{*}_{24} \hat{s}^{}_{25}
	\hat{s}^{*}_{35} \hat{s}^{}_{36} \\ + \hat{s}^{}_{14}
	\hat{s}^{*}_{24} c^{}_{25} \hat{s}^{}_{26} c^{}_{36} + c^{}_{14}
	\hat{s}^{}_{15} c^{}_{25} \hat{s}^{*}_{35}
	\hat{s}^{}_{36} \\
	+ c^{}_{14} \hat{s}^{}_{15} \hat{s}^{*}_{25} \hat{s}^{}_{26}
	c^{}_{36} - c^{}_{14} c^{}_{15} \hat{s}^{}_{16} c^{}_{26} c^{}_{36} \end{array} &
	\begin{array}{l} \hat{s}^{}_{24} \hat{s}^{*}_{34} c^{}_{35} \hat{s}^{}_{36}
	+ c^{}_{24} \hat{s}^{}_{25} \hat{s}^{*}_{35} \hat{s}^{}_{36} \\
	-c^{}_{24} c^{}_{25} \hat{s}^{}_{26} c^{}_{36} \end{array} &
	-c^{}_{34} c^{}_{35} \hat{s}^{}_{36} \cr \end{matrix} \right) \;.
\hspace{0.9cm}
\end{eqnarray}
The exact results in Eqs.~(10) and (11) have already been obtained in
Refs.~\cite{Xing:2011ur,Xing:2007zj,Xing:2019vks}
for the canonical (type-I) seesaw mechanism.
All the nine flavor mixing angles associated with $A^{}_1$, $B^{}_1$, $R^{}_1$ and
$S^{}_1$ are very small and can at most reach the level of ${\cal O}(0.1)$
\cite{Antusch:2006vwa,Fernandez-Martinez:2016lgt,Blennow:2016jkn,Wang:2021rsi},
as they describe the strength of active-sterile (or light-heavy) neutrino mixing
and thus are well constrained by current neutrino oscillation data and
precision measurements of various electroweak processes.
In fact, the smallness of $\theta^{}_{i4}$, $\theta^{}_{i5}$ and
$\theta^{}_{i6}$ (for $i=1,2,3$) implies that $A^{}_1$ and $B^{}_1$ are essentially
equal to the identity matrix $I$, and the magnitude of each element of
$R^{}_1$ and $S^{}_1$ is at most of ${\cal O}(0.1)$.
Similarly, the four $3\times 3$ submatrices of $L^{}_2$ can be expressed as
\begin{eqnarray}
A^{}_2 \hspace{-0.2cm} & = & \hspace{-0.2cm}
\left( \begin{matrix} c^{}_{17} c^{}_{18} c^{}_{19} & 0 & 0
	\cr\vspace{-0.4cm}\cr
	\begin{array}{l} -c^{}_{17} c^{}_{18} \hat{s}^{}_{19} \hat{s}^*_{29} -
	c^{}_{17} \hat{s}^{}_{18} \hat{s}^*_{28} c^{}_{29} \\
	-\hat{s}^{}_{17} \hat{s}^*_{27} c^{}_{28} c^{}_{29} \end{array} &
	c^{}_{27} c^{}_{28} c^{}_{29} & 0 \cr\vspace{-0.4cm}\cr
	\begin{array}{l} -c^{}_{17} c^{}_{18} \hat{s}^{}_{19} c^{}_{29} \hat{s}^*_{39}
	+ c^{}_{17} \hat{s}^{}_{18} \hat{s}^*_{28} \hat{s}^{}_{29} \hat{s}^*_{39} \\
	- c^{}_{17} \hat{s}^{}_{18} c^{}_{28} \hat{s}^*_{38} c^{}_{39} +
	\hat{s}^{}_{17} \hat{s}^*_{27} c^{}_{28} \hat{s}^{}_{29}
	\hat{s}^*_{39} \\
	+ \hat{s}^{}_{17} \hat{s}^*_{27} \hat{s}^{}_{28} \hat{s}^*_{38}
	c^{}_{39} - \hat{s}^{}_{17} c^{}_{27} \hat{s}^*_{37} c^{}_{38}
	c^{}_{39} \end{array} &
	\begin{array}{l} -c^{}_{27} c^{}_{28} \hat{s}^{}_{29} \hat{s}^*_{39} -
	c^{}_{27} \hat{s}^{}_{28} \hat{s}^*_{38} c^{}_{39} \\
	-\hat{s}^{}_{27} \hat{s}^*_{37} c^{}_{38} c^{}_{39} \end{array} &
	c^{}_{37} c^{}_{38} c^{}_{39} \cr \end{matrix} \right) \;,
\nonumber \\
B^{}_2 \hspace{-0.2cm} & = & \hspace{-0.2cm}
\left( \begin{matrix} c^{}_{17} c^{}_{27} c^{}_{37} & 0 & 0
	\cr\vspace{-0.4cm}\cr
	\begin{array}{l} -c^{}_{17} c^{}_{27} \hat{s}^{*}_{37} \hat{s}^{}_{38} -
	c^{}_{17} \hat{s}^{*}_{27} \hat{s}^{}_{28} c^{}_{38} \\
	-\hat{s}^{*}_{17} \hat{s}^{}_{18} c^{}_{28} c^{}_{38} \end{array} &
	c^{}_{18} c^{}_{28} c^{}_{38} & 0 \cr\vspace{-0.4cm}\cr
	\begin{array}{l} -c^{}_{17} c^{}_{27} \hat{s}^{*}_{37} c^{}_{38} \hat{s}^{}_{39}
	+ c^{}_{17} \hat{s}^{*}_{27} \hat{s}^{}_{28} \hat{s}^{*}_{38} \hat{s}^{}_{39} \\
	- c^{}_{17} \hat{s}^{*}_{27} c^{}_{28} \hat{s}^{}_{29} c^{}_{39} +
	\hat{s}^{*}_{17} \hat{s}^{}_{18} c^{}_{28} \hat{s}^{*}_{38}
	\hat{s}^{}_{39} \\
	+ \hat{s}^{*}_{17} \hat{s}^{}_{18} \hat{s}^{*}_{28} \hat{s}^{}_{29}
	c^{}_{39} - \hat{s}^{*}_{17} c^{}_{18} \hat{s}^{}_{19} c^{}_{29}
	c^{}_{39} \end{array} &
	\begin{array}{l} -c^{}_{18} c^{}_{28} \hat{s}^{*}_{38} \hat{s}^{}_{39} -
	c^{}_{18} \hat{s}^{*}_{28} \hat{s}^{}_{29} c^{}_{39} \\
	-\hat{s}^{*}_{18} \hat{s}^{}_{19} c^{}_{29} c^{}_{39} \end{array} &
	c^{}_{19} c^{}_{29} c^{}_{39} \cr \end{matrix} \right) \;;
\hspace{0.9cm}
\end{eqnarray}
and
\begin{eqnarray}
R^{}_2 \hspace{-0.2cm} & = & \hspace{-0.2cm}
\left( \begin{matrix} \hat{s}^*_{17} c^{}_{18} c^{}_{19} &
	\hat{s}^*_{18} c^{}_{19} & \hat{s}^*_{19} \cr\vspace{-0.4cm}\cr
	\begin{array}{l} -\hat{s}^*_{17} c^{}_{18} \hat{s}^{}_{19} \hat{s}^*_{29} -
	\hat{s}^*_{17} \hat{s}^{}_{18} \hat{s}^*_{28} c^{}_{29} \\
	+ c^{}_{17} \hat{s}^*_{27} c^{}_{28} c^{}_{29} \end{array} & -
	\hat{s}^*_{18} \hat{s}^{}_{19} \hat{s}^*_{29} + c^{}_{18}
	\hat{s}^*_{28} c^{}_{29} & c^{}_{19} \hat{s}^*_{29} \cr\vspace{-0.4cm}\cr
	\begin{array}{l} -\hat{s}^*_{17} c^{}_{18} \hat{s}^{}_{19} c^{}_{29}
	\hat{s}^*_{39} + \hat{s}^*_{17} \hat{s}^{}_{18} \hat{s}^*_{28}
	\hat{s}^{}_{29} \hat{s}^*_{39} \\ - \hat{s}^*_{17} \hat{s}^{}_{18}
	c^{}_{28} \hat{s}^*_{38} c^{}_{39} - c^{}_{17} \hat{s}^*_{27}
	c^{}_{28} \hat{s}^{}_{29}
	\hat{s}^*_{39} \\
	- c^{}_{17} \hat{s}^*_{27} \hat{s}^{}_{28} \hat{s}^*_{38}
	c^{}_{39} + c^{}_{17} c^{}_{27} \hat{s}^*_{37} c^{}_{38} c^{}_{39}
	\end{array} &
	\begin{array}{l} -\hat{s}^*_{18} \hat{s}^{}_{19} c^{}_{29} \hat{s}^*_{39}
	- c^{}_{18} \hat{s}^*_{28} \hat{s}^{}_{29} \hat{s}^*_{39} \\
	+c^{}_{18} c^{}_{28} \hat{s}^*_{38} c^{}_{39} \end{array} &
	c^{}_{19} c^{}_{29} \hat{s}^*_{39} \cr \end{matrix} \right) \;,
\nonumber \\
S^{}_2 \hspace{-0.2cm} & = & \hspace{-0.2cm}
\left( \begin{matrix} -\hat{s}^{}_{17} c^{}_{27} c^{}_{37}
& -\hat{s}^{}_{27} c^{}_{37} & -\hat{s}^{}_{37} \cr\vspace{-0.4cm}\cr
	\begin{array}{l} \hat{s}^{}_{17} c^{}_{27} \hat{s}^{*}_{37} \hat{s}^{}_{38}
	+ \hat{s}^{}_{17} \hat{s}^{*}_{27} \hat{s}^{}_{28} c^{}_{38} \\
	- c^{}_{17} \hat{s}^{}_{18} c^{}_{28} c^{}_{38} \end{array} &
	\hat{s}^{}_{27} \hat{s}^{*}_{37} \hat{s}^{}_{38} - c^{}_{27}
	\hat{s}^{}_{28} c^{}_{38} & -c^{}_{37} \hat{s}^{}_{38} \cr\vspace{-0.4cm}\cr
	\begin{array}{l} \hat{s}^{}_{17} c^{}_{27} \hat{s}^{*}_{37} c^{}_{38}
	\hat{s}^{}_{39} - \hat{s}^{}_{17} \hat{s}^{*}_{27} \hat{s}^{}_{28}
	\hat{s}^{*}_{38} \hat{s}^{}_{39} \\ + \hat{s}^{}_{17}
	\hat{s}^{*}_{27} c^{}_{28} \hat{s}^{}_{29} c^{}_{39} + c^{}_{17}
	\hat{s}^{}_{18} c^{}_{28} \hat{s}^{*}_{38}
	\hat{s}^{}_{39} \\
	+ c^{}_{17} \hat{s}^{}_{18} \hat{s}^{*}_{28} \hat{s}^{}_{29}
	c^{}_{39} - c^{}_{17} c^{}_{18} \hat{s}^{}_{19} c^{}_{29} c^{}_{39}
	\end{array} &
	\begin{array}{l} \hat{s}^{}_{27} \hat{s}^{*}_{37} c^{}_{38} \hat{s}^{}_{39}
	+ c^{}_{27} \hat{s}^{}_{28} \hat{s}^{*}_{38} \hat{s}^{}_{39} \\
	-c^{}_{27} c^{}_{28} \hat{s}^{}_{29} c^{}_{39} \end{array} &
	-c^{}_{37} c^{}_{38} \hat{s}^{}_{39} \cr \end{matrix} \right) \; .
\hspace{0.9cm}
\end{eqnarray}
One can see that the triangular matrices $A^{}_2$ and $B^{}_2$ have the same
forms as $A^{}_1$ and $B^{}_1$, and the former
can be directly obtained from the latter with the replacements
$\theta^{}_{i4} \to \theta^{}_{i7}$, $\theta^{}_{i5} \to \theta^{}_{i8}$,
$\theta^{}_{i6} \to \theta^{}_{i9}$ and
$\delta^{}_{i4} \to \delta^{}_{i7}$, $\delta^{}_{i5} \to \delta^{}_{i8}$,
$\delta^{}_{i6} \to \delta^{}_{i9}$ (for $i=1,2,3$). The same is true for
the forms of $R^{}_2$ and $S^{}_2$ as compared respectively
with those of $R^{}_1$ and $S^{}_1$. It is obvious that the flavor mixing
angles $\theta^{}_{i7}$, $\theta^{}_{i8}$ and $\theta^{}_{i9}$ are all
small because they describe the interplay between three active neutrinos
and three extra neutral fermions.
Finally, the four $3\times 3$ submatrices of $L^{}_3$ are
\begin{eqnarray}
A^{}_3 \hspace{-0.2cm} & = & \hspace{-0.2cm}
\left( \begin{matrix} c^{}_{47} c^{}_{48} c^{}_{49} & 0 & 0
	\cr\vspace{-0.4cm}\cr
	\begin{array}{l} -c^{}_{47} c^{}_{48} \hat{s}^{}_{49} \hat{s}^*_{59} -
	c^{}_{47} \hat{s}^{}_{48} \hat{s}^*_{58} c^{}_{59} \\
	-\hat{s}^{}_{47} \hat{s}^*_{57} c^{}_{58} c^{}_{59} \end{array} &
	c^{}_{57} c^{}_{58} c^{}_{59} & 0 \cr\vspace{-0.4cm}\cr
	\begin{array}{l} -c^{}_{47} c^{}_{48} \hat{s}^{}_{49} c^{}_{59} \hat{s}^*_{69}
	+ c^{}_{47} \hat{s}^{}_{48} \hat{s}^*_{58} \hat{s}^{}_{59} \hat{s}^*_{69} \\
	- c^{}_{47} \hat{s}^{}_{48} c^{}_{58} \hat{s}^*_{68} c^{}_{69} +
	\hat{s}^{}_{47} \hat{s}^*_{57} c^{}_{58} \hat{s}^{}_{59}
	\hat{s}^*_{69} \\
	+ \hat{s}^{}_{47} \hat{s}^*_{57} \hat{s}^{}_{58} \hat{s}^*_{68}
	c^{}_{69} - \hat{s}^{}_{47} c^{}_{57} \hat{s}^*_{67} c^{}_{68}
	c^{}_{69} \end{array} &
	\begin{array}{l} -c^{}_{57} c^{}_{58} \hat{s}^{}_{59} \hat{s}^*_{69} -
	c^{}_{57} \hat{s}^{}_{58} \hat{s}^*_{68} c^{}_{69} \\
	-\hat{s}^{}_{57} \hat{s}^*_{67} c^{}_{68} c^{}_{69} \end{array} &
	c^{}_{67} c^{}_{68} c^{}_{69} \cr \end{matrix} \right) \;,
\nonumber \\
B^{}_3 \hspace{-0.2cm} & = & \hspace{-0.2cm}
\left( \begin{matrix} c^{}_{47} c^{}_{57} c^{}_{67} & 0 & 0
	\cr\vspace{-0.4cm}\cr
	\begin{array}{l} -c^{}_{47} c^{}_{57} \hat{s}^{*}_{67} \hat{s}^{}_{68} -
	c^{}_{47} \hat{s}^{*}_{57} \hat{s}^{}_{58} c^{}_{68} \\
	-\hat{s}^{*}_{47} \hat{s}^{}_{48} c^{}_{58} c^{}_{68} \end{array} &
	c^{}_{48} c^{}_{58} c^{}_{68} & 0 \cr\vspace{-0.4cm}\cr
	\begin{array}{l} -c^{}_{47} c^{}_{57} \hat{s}^{*}_{67} c^{}_{68} \hat{s}^{}_{69}
	+ c^{}_{47} \hat{s}^{*}_{57} \hat{s}^{}_{58} \hat{s}^{*}_{68} \hat{s}^{}_{69} \\
	- c^{}_{47} \hat{s}^{*}_{57} c^{}_{58} \hat{s}^{}_{59} c^{}_{69} +
	\hat{s}^{*}_{47} \hat{s}^{}_{48} c^{}_{58} \hat{s}^{*}_{68}
	\hat{s}^{}_{69} \\
	+ \hat{s}^{*}_{47} \hat{s}^{}_{48} \hat{s}^{*}_{58} \hat{s}^{}_{59}
	c^{}_{69} - \hat{s}^{*}_{47} c^{}_{48} \hat{s}^{}_{49} c^{}_{59}
	c^{}_{69} \end{array} &
	\begin{array}{l} -c^{}_{48} c^{}_{58} \hat{s}^{*}_{68} \hat{s}^{}_{69} -
	c^{}_{48} \hat{s}^{*}_{58} \hat{s}^{}_{59} c^{}_{69} \\
	-\hat{s}^{*}_{48} \hat{s}^{}_{49} c^{}_{59} c^{}_{69} \end{array} &
	c^{}_{49} c^{}_{59} c^{}_{69} \cr \end{matrix} \right) \;;
\hspace{0.9cm}
\end{eqnarray}
and
\begin{eqnarray}
R^{}_3 \hspace{-0.2cm} & = & \hspace{-0.2cm}
\left( \begin{matrix} \hat{s}^*_{47} c^{}_{48} c^{}_{49} &
	\hat{s}^*_{48} c^{}_{49} & \hat{s}^*_{49} \cr\vspace{-0.4cm}\cr
	\begin{array}{l} -\hat{s}^*_{47} c^{}_{48} \hat{s}^{}_{49} \hat{s}^*_{59} -
	\hat{s}^*_{47} \hat{s}^{}_{48} \hat{s}^*_{58} c^{}_{59} \\
	+ c^{}_{47} \hat{s}^*_{57} c^{}_{58} c^{}_{59} \end{array} & -
	\hat{s}^*_{48} \hat{s}^{}_{49} \hat{s}^*_{59} + c^{}_{48}
	\hat{s}^*_{58} c^{}_{59} & c^{}_{49} \hat{s}^*_{59} \cr\vspace{-0.4cm}\cr
	\begin{array}{l} -\hat{s}^*_{47} c^{}_{48} \hat{s}^{}_{49} c^{}_{59}
	\hat{s}^*_{69} + \hat{s}^*_{47} \hat{s}^{}_{48} \hat{s}^*_{58}
	\hat{s}^{}_{59} \hat{s}^*_{69} \\ - \hat{s}^*_{47} \hat{s}^{}_{48}
	c^{}_{58} \hat{s}^*_{68} c^{}_{69} - c^{}_{47} \hat{s}^*_{57}
	c^{}_{58} \hat{s}^{}_{59}
	\hat{s}^*_{69} \\
	- c^{}_{47} \hat{s}^*_{57} \hat{s}^{}_{58} \hat{s}^*_{68}
	c^{}_{69} + c^{}_{47} c^{}_{57} \hat{s}^*_{67} c^{}_{68} c^{}_{69}
	\end{array} &
	\begin{array}{l} -\hat{s}^*_{48} \hat{s}^{}_{49} c^{}_{59} \hat{s}^*_{69}
	- c^{}_{48} \hat{s}^*_{58} \hat{s}^{}_{59} \hat{s}^*_{69} \\
	+c^{}_{48} c^{}_{58} \hat{s}^*_{68} c^{}_{69} \end{array} &
	c^{}_{49} c^{}_{59} \hat{s}^*_{69} \cr \end{matrix} \right) \;,
\nonumber \\
S^{}_3 \hspace{-0.2cm} & = & \hspace{-0.2cm}
\left( \begin{matrix} -\hat{s}^{}_{47} c^{}_{57} c^{}_{67} &
	-\hat{s}^{}_{57} c^{}_{67} & -\hat{s}^{}_{67} \cr\vspace{-0.4cm}\cr
	\begin{array}{l} \hat{s}^{}_{47} c^{}_{57} \hat{s}^{*}_{67} \hat{s}^{}_{68}
	+ \hat{s}^{}_{47} \hat{s}^{*}_{57} \hat{s}^{}_{58} c^{}_{68} \\
	- c^{}_{47} \hat{s}^{}_{48} c^{}_{58} c^{}_{68} \end{array} &
	\hat{s}^{}_{57} \hat{s}^{*}_{67} \hat{s}^{}_{68} - c^{}_{57}
	\hat{s}^{}_{58} c^{}_{68} & -c^{}_{67} \hat{s}^{}_{68} \cr\vspace{-0.4cm}\cr
	\begin{array}{l} \hat{s}^{}_{47} c^{}_{57} \hat{s}^{*}_{67} c^{}_{68}
	\hat{s}^{}_{69} - \hat{s}^{}_{47} \hat{s}^{*}_{57} \hat{s}^{}_{58}
	\hat{s}^{*}_{68} \hat{s}^{}_{69} \\ + \hat{s}^{}_{47}
	\hat{s}^{*}_{57} c^{}_{58} \hat{s}^{}_{59} c^{}_{69} + c^{}_{47}
	\hat{s}^{}_{48} c^{}_{58} \hat{s}^{*}_{68}
	\hat{s}^{}_{69} \\
	+ c^{}_{47} \hat{s}^{}_{48} \hat{s}^{*}_{58} \hat{s}^{}_{59}
	c^{}_{69} - c^{}_{47} c^{}_{48} \hat{s}^{}_{49} c^{}_{59} c^{}_{69}
	\end{array} &
	\begin{array}{l} \hat{s}^{}_{57} \hat{s}^{*}_{67} c^{}_{68} \hat{s}^{}_{69}
	+ c^{}_{57} \hat{s}^{}_{58} \hat{s}^{*}_{68} \hat{s}^{}_{69} \\
	-c^{}_{57} c^{}_{58} \hat{s}^{}_{59} c^{}_{69} \end{array} &
	-c^{}_{67} c^{}_{68} \hat{s}^{}_{69} \cr \end{matrix} \right) \;.
\hspace{0.9cm}
\end{eqnarray}
Different from the eighteen active-sterile flavor mixing angles $\theta^{}_{1j}$,
$\theta^{}_{2j}$ and $\theta^{}_{3j}$ (for $j=4,5,\cdots,9$), which are all expected
to be strongly suppressed in magnitude, the nine sterile-sterile flavor mixing angles
$\theta^{}_{4j}$, $\theta^{}_{5j}$ and $\theta^{}_{6j}$ (for $j=7,8,9$) are completely
unconstrained. Given the very fact that the right-handed neutrino fields and the
extra neutral fermion fields are both hypothetical, one may naively conjecture that
these two sterile sectors might be essentially disconnected and thus their interplay
might be very weak. Here we only focus our attention on the issue of active-sterile
flavor mixing. Without loss of any generality, we have chosen the basis in which the
flavor eigenstates of three charged leptons are identical with their mass eigenstates
throughout this work.

After Eqs.~(6), (7) and (8) are taken into account, the $9\times 9$ active-sterile
flavor mixing matrix $\cal U$ can be explicitly written as
\begin{eqnarray}
{\cal U} = \left( \begin{matrix} A^{}_2 A^{}_1 U^{}_0 & A^{}_2 R^{}_1 & R^{}_2 \cr
\big(R^{}_3 S^{}_2 A^{}_1 + A^{}_3 U^{\prime}_0 S^{}_1\big) U^{}_0
& R^{}_3 S^{}_2 R^{}_1 + A^{}_3 U^{\prime}_0 B^{}_1 & R^{}_3 B^{}_2 \cr
S^{}_0 \big(B^{}_3 S^{}_2 A^{}_1 + S^{}_3 U^{\prime}_0 S^{}_1\big) U^{}_0
& ~ S^{}_0 \big(B^{}_3 S^{}_2 R^{}_1 + S^{}_3 U^{\prime}_0 B^{}_1\big) ~
& S^{}_0 B^{}_3 B^{}_2 \end{matrix} \right) \; .
\end{eqnarray}
The unitarity of $\cal U$ (i.e., ${\cal U} {\cal U}^\dagger = {\cal U}^\dagger
{\cal U} = I^{}_{9\times 9}$ with $I^{}_{9 \times 9}$ being the $9\times 9$
identity matrix) allows us to obtain a number of constraint equations, as listed
in appendix~\ref{appendix A}. In the chosen flavor basis the key role of $\cal U$ is to
transform the flavor eigenstates of three active neutrinos, three sterile
neutrinos and three extra neutral fermions shown in Eq.~(2)
into their mass eigenstates; namely,
\begin{eqnarray}
{\cal U}^\dagger \left( \begin{matrix} 0 & M^{}_{\rm D} & \varepsilon
\cr M^T_{\rm D} & M^{}_{\rm R} & M^{}_S \cr \varepsilon^{T} & M^T_S &
\mu \cr \end{matrix} \right) {\cal U}^* =
\left( \begin{matrix} D^{}_{\nu} & 0 & 0 \cr 0 & D^{}_N & 0 \cr
0 & 0 & D^{}_S \end{matrix} \right) \; ,
\end{eqnarray}
where $D^{}_\nu = \{m^{}_1, m^{}_2, m^{}_3\}$,
$D^{}_N = \{M^{}_1, M^{}_2, M^{}_3\}$ and
$D^{}_S = \{M^{\prime}_1, M^{\prime}_2, M^{\prime}_3\}$ with $m^{}_i$,
$M^{}_i$ and $M^{\prime}_i$ being the respective masses
of the active neutrinos $\nu^{}_i$, sterile neutrinos $N^{}_i$ and
extra neutral fermions $N^{\prime}_i$ (for $i=1,2,3$). We are therefore left with
\begin{eqnarray}
\left( \begin{matrix} \nu^{}_e \cr \nu^{}_\mu \cr \nu^{}_\tau \end{matrix}
\right)_{\hspace{-0.1cm} \rm L}
= A^{}_2 A^{}_1 U^{}_0 \left( \begin{matrix} \nu^{}_1 \cr \nu^{}_2 \cr \nu^{}_3
\end{matrix} \right)_{\hspace{-0.1cm} \rm L} + A^{}_2 R^{}_1
\left( \begin{matrix} N^{}_1 \cr N^{}_2 \cr N^{}_3
\end{matrix} \right)_{\hspace{-0.1cm} \rm L} + R^{}_2 \left( \begin{matrix}
N^{\prime}_1 \cr N^{\prime}_2 \cr N^{\prime}_3 \end{matrix}
\right)_{\hspace{-0.1cm} \rm L} \; .
\end{eqnarray}
The weak charged-current interactions of three active neutrinos, three sterile
neutrinos and three extra neutral fermions turn out to be
\begin{eqnarray}
-{\cal L}^{}_{\rm cc} = \frac{g}{\sqrt{2}} \
\overline{\left(e ~~ \mu ~~ \tau\right)^{}_{\rm L}} \ \gamma^\mu \left[ U
\left(\begin{matrix} \nu^{}_{1} \cr \nu^{}_{2} \cr
\nu^{}_{3}\end{matrix}\right)^{}_{\hspace{-0.1cm} \rm L} +
R \left( \begin{matrix} N^{}_1 \cr N^{}_2 \cr N^{}_3
\end{matrix} \right)_{\hspace{-0.1cm} \rm L} + R^{\prime} \left( \begin{matrix}
N^{\prime}_1 \cr N^{\prime}_2 \cr N^{\prime}_3 \end{matrix}
\right)_{\hspace{-0.1cm} \rm L} \right] W^-_\mu + {\rm h.c.} \; ,
\end{eqnarray}
where $U \equiv A^{}_2 A^{}_1 U^{}_0$ represents the effective PMNS matrix,
$R \equiv A^{}_2 R^{}_1$ and $R^\prime \equiv R^{}_2$ characterize the
contributions of new degrees of freedom to ${\cal L}^{}_{\rm cc}$.
Once the extra neutral fermions are switched off (i.e., $A^{}_2 = I$
and $R^{}_2 = 0$), Eq.~(19) will be reduced to the canonical seesaw case
\cite{Xing:2011ur,Xing:2007zj,Xing:2019vks}.

Note that the $3\times 3$ PMNS matrix $U$ is not exactly unitary, and its
slight deviation from the unitarity limit can be clearly seen from
\begin{eqnarray}
U U^{\dagger} = A^{}_2 A^{}_1 A^{\dagger}_1 A^{\dagger}_2
= I - R R^{\dagger} - R^{\prime} R^{\prime \dagger} \; ,
\end{eqnarray}
where the relation in Eq.~(A.1) has been used. This issue will be further
discussed in section 4.

Note also that the full parametrization of the $9\times 9$ unitary
flavor mixing matrix $\cal U$ obtained in Eq.~(16) is a quite general result,
and it is actually not subject to the seesaw scenario and the corresponding
$9\times 9$ neutrino mass matrix considered in Eqs.~(1) and
(2). For example, the texture zeros in the original neutrino mass matrix
will help establish some correlations between the mass and flavor mixing
parameters, including the exact seesaw formulas. This point will be clearly
seen later on in the inverse and linear seesaw mechanisms.

\section{Exact and approximate seesaw formulas}

\subsection{The inverse seesaw scenario}

As for the inverse seesaw scenario with $M^{}_{\rm R} =0$ and $\varepsilon =0$,
the corresponding neutrino mass matrix takes the well-known Fritzsch
texture \cite{Fritzsch:1977vd}
\begin{eqnarray}
{\cal F} = \left( \begin{matrix} 0 & M^{}_{\rm D} & 0 \cr
M^{T}_{\rm D} & 0 & M^{}_S \cr 0 & M^{T}_S & \mu \end{matrix} \right)
= {\cal U} \left( \begin{matrix} D^{}_{\nu} & 0 & 0 \cr 0 & D^{}_N & 0 \cr
0 & 0 & D^{}_S \end{matrix} \right) {\cal U}^T \; .
\end{eqnarray}
The texture zero ${\cal F}^{}_{11} = 0$ allows us to obtain the {\it exact}
inverse seesaw formula
\begin{eqnarray}
U D^{}_\nu U^T + R D^{}_N R + R^{\prime} D^{}_S R^{\prime T} = 0 \; ,
\end{eqnarray}
where $U \equiv A^{}_2 A^{}_1 U^{}_0$, $R \equiv A^{}_2 R^{}_1$ and
$R^\prime \equiv R^{}_2$ have been defined below Eq.~(19), and they are
correlated with one another through
$UU^\dagger + RR^\dagger + R^\prime R^{\prime \dagger} = I$ as can be
seen from Eq.~(20). If the extra neutral fermion sector is switched off
(i.e., $A^{}_2 = I$ and $R^\prime = 0$), one will reproduce the exact
seesaw formula of the canonical seesaw mechanism
\cite{Xing:2011ur,Xing:2007zj,Xing:2019vks}. Moreover, the texture zeros
${\cal F}^{}_{22} = 0$ and ${\cal F}^{}_{13} = {\cal F}^{}_{31} = 0$
lead us to the constraint equations
\begin{eqnarray}
{\rm (a)}: \hspace{-0.2cm} & & \hspace{-0.2cm}
\big( R^{}_3 S^{}_2 A^{}_1 U^{}_0 + A^{}_3 U^{\prime}_0 S^{}_1 U^{}_0 \big)
D^{}_\nu \big( U^T_{0} A^T_1 S^T_2 R^T_3 + U^T_{0} S^T_1 U^{\prime T}_0
A^T_3 \big)
\nonumber \\
\hspace{-0.2cm} & & \hspace{-0.2cm}
+ \big( R^{}_3 S^{}_2 R^{}_1 + A^{}_3 U^{\prime}_0 B^{}_1 \big) D^{}_N
\big( R^T_1 S^T_2 R^T_3 + B^T_1 U^{\prime T}_0 A^T_3 \big)
+ R^{}_3 B^{}_2 D^{}_S B^T_2 R^T_3 = 0 \; , \hspace{0.5cm}
\nonumber \\
{\rm (b)}: \hspace{-0.2cm} & & \hspace{-0.2cm}
\big( S^{}_0 B^{}_3 S^{}_2 A^{}_1 U^{}_0 + S^{}_0 S^{}_3 U^{\prime}_0
S^{}_1 U^{}_0 \big) D^{}_\nu U^T_0 A^T_1 A^T_2
\nonumber \\
\hspace{-0.2cm} & & \hspace{-0.2cm}
+ \big( S^{}_0 B^{}_3 S^{}_2 R^{}_1 + S^{}_0 S^{}_3 U^{\prime}_0 B^{}_1 \big)
D^{}_N R^T_1 A^T_2 + S^{}_0 B^{}_3 B^{}_2 D^{}_S R^{T}_2 = 0 \; .
\end{eqnarray}
These two equations, together with Eq.~(22), characterize the correlations between
the mass and flavor mixing parameters in this seesaw scenario.

We proceed to derive the approximate but more instructive inverse seesaw formula
given in Eq.~(3) from Eqs.~(22) and (23). To this end, we take into account
the fact that the interplay between three active neutrinos and those hypothetical
degrees of freedom must be strongly suppressed in magnitude. In other words, the
active-sterile flavor mixing angles $\theta^{}_{1j}$, $\theta^{}_{2j}$ and
$\theta^{}_{3j}$ (for $j=4, 5, \cdots, 9$) are all very small, and thus the
eight $3\times 3$ matrices $A^{}_{1,2}$, $B^{}_{1,2}$, $R^{}_{1,2}$ and $S^{}_{1,2}$
in Eqs.~(10)---(13) can be reliably simplified to the forms listed in
appendix~\ref{appendix B}. As very good approximations, Eqs.~(22) and (23) become
\begin{eqnarray}
U^{}_0 D^{}_\nu U^T_0 + R^{}_1 D^{}_N R^T_1 + R^{}_2 D^{}_S R^T_2 \simeq 0 \; ,
\end{eqnarray}
and
\begin{eqnarray}
\hspace{-0.2cm} & & \hspace{-0.2cm}
A^{}_3 U^{\prime}_0 D^{}_N U^{\prime T}_0 A^T_3 + R^{}_3 D^{}_S R^T_3 \simeq 0 \; ,
\hspace{1cm}
\nonumber \\
\hspace{-0.2cm} & & \hspace{-0.2cm}
S^{}_3 U^{\prime}_0 D^{}_N R^T_1 + B^{}_3 D^{}_S R^T_2 \simeq 0 \; .
\end{eqnarray}
Since $D^{}_\nu$ is strongly suppressed in magnitude as compared with $D^{}_N$ and 
$D^{}_S$, the sum of the second and third terms on the left-hand side of Eq.~(24) 
should be as small as the first term and possess the opposite sign. 
Moreover, $D^{}_N \simeq -D^{}_S$ is expected to hold due to
the smallness of $\mu$, as one can see when solving the eigenvalue equation
of $\cal F$. This observation implies
that $R^{}_1 \simeq R^{}_2$ is a good approximation. Then
$A^{}_3 U^{\prime}_0 \simeq -R^{}_3$ and $S^{}_3 U^{\prime}_0 \simeq B^{}_3$
can be derived from Eq.~(25), and their opposite signs will be explained later.
In this case the effective $3\times 3$ Majorana
neutrino mass matrix of three active neutrinos can be defined as
\begin{eqnarray}
M^{}_\nu \equiv U^{}_0 D^{}_\nu U^T_0 \simeq
-R^{}_1 D^{}_N R^T_1 - R^{}_2 D^{}_S R^T_2 \; ,
\end{eqnarray}
where Eq.~(24) has been used. To see how $M^{}_\nu$ is related to $M^{}_{\rm D}$,
$M^{}_S$ and $\mu$, we start from Eq.~(21) and obtain the following results:
\begin{eqnarray}
M^{}_{\rm D} \hspace{-0.2cm} & = & \hspace{-0.2cm}
R^{}_1 D^{}_N \left( R^T_1 S^T_2 R^T_3 + U^{\prime T}_{0} A^T_3 \right) +
R^{}_2 D^{}_S R^T_3
\nonumber \\
\hspace{-0.2cm} & \simeq & \hspace{-0.2cm}
R^{}_1 D^{}_N  U^{\prime T}_{0} A^T_3 + R^{}_2 D^{}_S R^T_3
\nonumber \\
\hspace{-0.2cm} & \simeq & \hspace{-0.2cm}
R^{}_1  D^{}_N \big( U^{\prime T}_{0} A^T_3 - R^T_3\big)
\nonumber \\
\hspace{-0.2cm} & \simeq & \hspace{-0.2cm}
2R^{}_1 D^{}_N  U^{\prime T}_{0} A^T_3 \; ,
\nonumber \\
M^{}_S \hspace{-0.2cm} & = & \hspace{-0.2cm}
\big(R^{}_3 S^{}_2 R^{}_1 + A^{}_3 U^{\prime}_0 \big) D^{}_N
\left( R^T_1 S^T_2 B^T_3 S^T_{0} + U^{\prime T}_0 S^T_3 S^T_{0}\right)
+ R^{}_3 D^{}_S B^T_3 S^T_0
\nonumber \\
\hspace{-0.2cm} & \simeq & \hspace{-0.2cm}
A^{}_3 U^{\prime}_0 D^{}_N U^{\prime T}_0 S^T_3 S^T_0 + R^{}_3 D^{}_S B^T_3 S^T_0
\nonumber \\	
\hspace{-0.2cm} & \simeq & \hspace{-0.2cm}
\big( A^{}_3 U^{\prime}_0 - R^{}_3 \big) D^{}_N U^{\prime T}_0 S^T_3 S^T_0
\nonumber \\
\hspace{-0.2cm} & \simeq & \hspace{-0.2cm}
2A^{}_3 U^{\prime}_0 D^{}_N U^{\prime T}_0 S^T_3 S^T_0 \; ,
\nonumber \\
\mu \hspace{-0.2cm} & = & \hspace{-0.2cm}
\big(S^{}_0 B^{}_3 S^{}_2 R^{}_1 + S^{}_0 S^{}_3 U^{\prime}_0 B^{}_1 \big)
D^{}_N \left( R^T_2 S^T_2 B^T_3 S^T_{0} + B^T_1 U^{\prime T}_0 S^T_3 S^T_{0}
\right) + S^{}_0 B^{}_3 B^{}_2 D^{}_S B^T_2 B^T_3 S^T_0 \hspace{0.5cm}
\nonumber \\
\hspace{-0.2cm} & \simeq & \hspace{-0.2cm}	
S^{}_0 S^{}_3 U^{\prime}_0 D^{}_N U^{\prime T}_0 S^T_3 S^T_{0}
+ S^{}_0 B^{}_3 D^{}_S B^T_3 S^T_0
\nonumber \\
\hspace{-0.2cm} & \simeq & \hspace{-0.2cm}	
S^{}_0 B^{}_3 \big(B^{-1}_3 S^{}_3 U^{\prime}_0 D^{}_N D^{-1}_N
D^{}_N U^{\prime T}_0 S^{T}_3 B^{T-1}_3 + D^{}_S \big) B^T_3 S^T_0
\nonumber \\
\hspace{-0.2cm} & \simeq & \hspace{-0.2cm}	
S^{}_0 B^{}_3 \big(-R^{-1}_1 R^{}_2 D^{}_S R^{T}_2 R^{T-1}_1 - D^{}_N \big) B^T_3 S^T_0 \; .
\end{eqnarray}
It is known that a natural TeV-scale inverse seesaw scenario requires
$M^{}_{\rm D} \sim {\cal O}(1)$ GeV, $M^{}_S \sim {\cal O}(1)$ TeV and
$\mu \sim {\cal O}(1)$ keV (see, e.g., Ref.~\cite{Malinsky:2009df}).
Hence the two terms in $M^{}_{\rm D}$ and $M^{}_S$ should be of the same sign
to avoid a significant cancellation. That is why we have chosen
$A^{}_3 U^{\prime}_0 \simeq -R^{}_3$ below Eq.~(25).
On the other hand, the two terms in $\mu$ should nearly cancel each other,
implying that $S^{}_3 U^{\prime}_0 \simeq B^{}_3$ is a reasonable choice.
Substituting the first equation in Eq.~(25) into Eq.~(26) and then
making use of the approximate expressions obtained in Eq.~(27), we simply arrive at
\begin{eqnarray}
M^{}_\nu \hspace{-0.2cm} & \simeq & \hspace{-0.2cm}
-R^{}_1 D^{}_N R^T_1-R^{}_2 D^{}_S R^T_2
\nonumber \\
\hspace{-0.2cm} & \simeq & \hspace{-0.2cm}
R^{}_1 \left( -D^{}_N -R^{-1}_1 R^{}_2 D^{}_S R^T_2 R^{T-1}_1\right) R^{T}_1
\nonumber \\
\hspace{-0.2cm} & \simeq & \hspace{-0.2cm}
R^{}_1 U^{\prime -1}_0 S^{-1}_3 B^{}_3 \left( -D^{}_N -R^{-1}_1 R^{}_2 D^{}_S R^T_2 R^{T-1}_1\right) B^{T}_3 S^{T-1}_3 \left(U^{\prime T}_0\right)^{-1} R^{T}_1
\nonumber \\
\hspace{-0.2cm} & \simeq & \hspace{-0.2cm}
\Big[R^{}_1 U^{\prime -1}_0 S^{-1}_3 S^{-1}_0\Big] \mu
\Big[R^{}_1 U^{\prime -1}_0 S^{-1}_3 S^{-1}_0\Big]^T
\nonumber \\
\hspace{-0.2cm} & \simeq & \hspace{-0.2cm}
\Big[2 R^{}_1 D^{}_N  U^{\prime T}_{0} A^T_3
\big(2 S^{}_0 S^{}_3 U^\prime_0 D^{}_N U^{\prime T}_0 A^T_3\big)^{-1}\Big] \mu
\Big[\big(2 A^{}_3 U^{\prime}_0 D^{}_N U^{\prime T}_0 S^T_3 S^T_0\big)^{-1}
\big(2 R^{}_1 D^{}_N  U^{\prime T}_{0} A^T_3 \big)^T\Big]
\hspace{0.5cm}
\nonumber \\
\hspace{-0.2cm} & \simeq & \hspace{-0.2cm}
M^{}_{\rm D} \left(M^{T}_S\right)^{-1} \mu \left(M^{}_S\right)^{-1} M^{T}_{\rm D} \; .
\end{eqnarray}
This leading-order result is just the well-known inverse seesaw
formula given in Eq.~(3). While such an approximate expression may be more instructive
in some explicit model-building exercises, our exact inverse seesaw formula and full
parametrization of the $9\times 9$ active-sterile flavor mixing matrix will be
more useful for a generic study of the inverse seesaw picture.

At this point it is worth remarking that we have only paid attention to the
leading-order effects when defining the effective Majorana neutrino
mass matrix of three active neutrinos in Eq.~(26) and reproducing the
inverse seesaw formula for it in Eqs.~(27) and (28). The next-to-leading-order
corrections to the leading-order inverse seesaw formula have been discussed
at the tree level \cite{Hettmansperger:2011bt}. We find that it will be a messy
and tangled business to include such next-to-leading-order effects into
our approach, because both our generic Euler-like parametrization of the
$9\times 9$ active-sterile flavor mixing matrix and our exact inverse seesaw
formula are actually independent of the sub-leading effects of this kind.
This is also true of our discussions about the non-unitary effects on flavor
mixing and CP violation based on our parametrization in section~\ref{section:4}.

\subsection{The linear seesaw scenario}

Since $M^{}_{\rm R} = 0$ and $\mu = 0$ are taken for the linear seesaw scenario,
one may reconstruct the corresponding traceless neutrino mass matrix in the
following way:
\begin{eqnarray}
{\cal F}^\prime = \left( \begin{matrix} 0 & M^{}_{\rm D} & \varepsilon \cr
M^{T}_{\rm D} & 0 & M^{}_S \cr \varepsilon^{T}_{} & M^{T}_S & 0
\end{matrix} \right) = {\cal U} \left( \begin{matrix}
D^{}_{\nu} & 0 & 0 \cr 0 & {D^{}_N} & 0 \cr 0 & 0 & {D^{}_S} \end{matrix}
\right) {\cal U^{T}_{}} \; .
\end{eqnarray}
In this case the texture zero ${\cal F}^\prime_{11} = 0$ leads us to the
{\it exact} linear seesaw formula
\begin{eqnarray}
U D^{}_\nu U^T + R D^{}_N R + R^{\prime} D^{}_S R^{\prime T} = 0 \; ,
\end{eqnarray}
where $U \equiv A^{}_2 A^{}_1 U^{}_0$, $R \equiv A^{}_2 R^{}_1$ and
$R^\prime \equiv R^{}_2$ are defined and
$UU^\dagger + RR^\dagger + R^\prime R^{\prime \dagger} = I$ holds. One can immediately
see that Eq.~(30) is formally the same as Eq.~(22) obtained in the inverse seesaw
case, although their physical contexts are somewhat different. This interesting
observation tells us that such an {\it exact} seesaw formula is actually universal
and valid for the more general seesaw scenario described by the mass terms in
Eqs.~(1) and (2).

Taking account of the texture zeros ${\cal F}^{\prime}_{22} = {\cal F}^\prime_{33} = 0$
in the linear seesaw scenario, we arrive at two further constraint equations
\begin{eqnarray}
{\rm (a)}: \hspace{-0.2cm} & & \hspace{-0.2cm}	
\big( R^{}_{3} S^{}_{2} A^{}_{1} U^{}_0 + A^{}_{3} U^{\prime}_0 S^{}_{1} U^{}_0 \big)
D^{}_\nu \big( U^T_{0} A^T_{1} S^T_{2} R^T_{3}
+ U^T_{0} S^T_{1} U^{\prime T}_0 A^T_{3} \big)
\nonumber \\
\hspace{-0.2cm} & & \hspace{-0.2cm}	
+ \big( R^{}_{3} S^{}_{2} R^{}_{1} + A^{}_{3} U^{\prime}_0 B^{}_{1}\big)
D^{}_N \big(R^T_{1} S^T_{2} R^T_{3} + B^T_{1} U^{\prime T}_0 A^T_{3}\big)
+ R^{}_{3} B^{}_{2} D^{}_S B^T_{2} R^T_{3} = 0 \; ,
\nonumber \\
{\rm (b)}: \hspace{-0.2cm} & & \hspace{-0.2cm}	
\big( S^{}_0 B^{}_{3} S^{}_{2} A^{}_{1} U^{}_0 + S^{}_0 S^{}_{3} U^{\prime}_0 S^{}_{1}
U^{}_0 \big) D^{}_\nu \big( U^T_{0} A^T_{1} S^T_{2} B^T_{3} S^{T}_0 + U^T_{0} S^T_{1}
U^{\prime T}_0 S^T_{3} S^{T}_0 \big)
\nonumber \\
\hspace{-0.2cm} & & \hspace{-0.2cm}	
+ \big( S^{}_0 B^{}_{3} S^{}_{2} R^{}_{1} + S^{}_0 S^{}_{3} U^{\prime}_0 B^{}_{1}
\big) D^{}_N \big( B^{T}_{1} U^{\prime T}_0 S^{T}_{3} S^{T}_0 + R^{T}_{1} S^{T}_{2}
B^{T}_{3} S^{T}_0 \big)
+ S^{}_0 B^{}_{3} D^{}_S B^T_{3} S^{T}_0 = 0 \; . \hspace{0.5cm}
\end{eqnarray}
Once the approximations made in Eqs.~(B1)---(B4) are taken into consideration,
Eqs.~(30) and (31) can be easily simplified to
\allowdisplaybreaks[4]
\begin{eqnarray}
U^{}_0 D^{}_\nu U^T_0 + R^{}_{1} D^{}_N R^T_{1} + R^{}_{2} D^{}_S R^T_{2} \simeq 0 \; ,
\end{eqnarray}
and
\begin{eqnarray}
\hspace{-0.2cm} & & \hspace{-0.2cm}
A^{}_{3} U^{\prime}_0 D^{}_N U^{\prime T}_0 A^T_{3} + R^{}_{3} D^{}_S R^T_{3} \simeq 0 \; ,
\nonumber \\
\hspace{-0.2cm} & & \hspace{-0.2cm}
S^{}_{3} U^{\prime}_0 D^{}_N U^{\prime T}_0 S^{T}_{3} + B^{}_{3} D^{}_S B^T_{3} \simeq 0 \; .
\hspace{1.1cm}
\end{eqnarray}
As in the inverse seesaw scenario, we similarly have $D^{}_N \simeq -D^{}_S$ and
$R^{}_1 \simeq R^{}_2$ in the linear seesaw scenario. The relations
$A^{}_3 U^{\prime}_0 \simeq -R^{}_3$ and $S^{}_3 U^{\prime}_0 \simeq B^{}_3$ are expected
to remain valid. The reason why these two approximate relations have the
opposite signs is also that the terms in $M^{}_{\rm D}$ and $M^{}_S$ should avoid
significant cancellations while $\varepsilon$ should be strongly suppressed in magnitude.

Now let us use Eq.~(32) to define the effective $3\times 3$
Majorana neutrino mass matrix of three active neutrinos:
\begin{eqnarray}
M^{}_\nu \equiv U^{}_0 D^{}_\nu U^T_0 \simeq
-R^{}_{1} D^{}_N R^T_{1}-R^{}_{2} D^{}_S R^T_{2} \; .
\end{eqnarray}
Starting from Eq.~(29) and taking into account the approximate relations obtained in
and below Eq.~(33), we have
\begin{eqnarray}
M^{}_{\rm D} \hspace{-0.2cm} & = & \hspace{-0.2cm}
R^{}_{1} D^{}_N \big( R^T_{1} S^T_{2} R^T_{3} + U^{\prime T}_0 A^{T}_{3} \big)
+ R^{}_{2} D^{}_S R^T_{3}
\nonumber \\
\hspace{-0.2cm} & \simeq & \hspace{-0.2cm}
2R^{}_{1} D^{}_N U^{\prime T}_0 A^{T}_{3} \; ,
\nonumber \\
M^{}_S \hspace{-0.2cm} & = & \hspace{-0.2cm}
\big(R^{}_{3} S^{}_{2} R^{}_{1} + A^{}_{3} U^{\prime}_0\big) D^{}_N
\big( R^T_{1} S^T_{2} B^T_{3} S^T_{0} + U^{\prime T}_0 S^T_{3} S^T_{0}\big)
+ R^{}_{3} D^{}_S B^T_{3} S^T_0
\hspace{0.5cm}
\nonumber \\
\hspace{-0.2cm} & \simeq & \hspace{-0.2cm}
2 A^{}_{3} U^{\prime}_0 D^{}_N U^{\prime T}_0 S^T_{3} S^T_{0}
\nonumber \\
\varepsilon \hspace{-0.2cm} & = & \hspace{-0.2cm}
R^{}_{1} D^{}_N \big( R^T_{1} S^T_{2} B^T_{3} S^T_{0} +
U^{\prime T}_0 S^T_{3} S^T_{0} \big)
+ R^{}_{2} D^{}_S B^T_{3} S^T_0
\nonumber \\
\hspace{-0.2cm} & \simeq & \hspace{-0.2cm}
R^{}_{1} D^{}_N U^{\prime T}_0 S^T_{3} S^T_{0} + R^{}_{2} D^{}_S B^T_{3} S^T_0 \; .
\end{eqnarray}
It is then easy to arrive at
\begin{eqnarray}
R^{}_{1} D^{}_N R^T_{1} - R^{}_{1} D^{}_N R^T_{2}
\hspace{-0.2cm} & \simeq & \hspace{-0.2cm}
R^{}_{1} U^{\prime -1}_0 S^{-1}_3 \left( S^{}_3 U^{\prime}_0 D^{}_N R^T_{1}
+ B^{}_3 D^{}_S R^T_{2} \right)
\nonumber \\
\hspace{-0.2cm} & \simeq & \hspace{-0.2cm}
\left(2R^{}_{1} D^{}_N U^{\prime T}_0 A^{T}_{3}\right)
\left( 2 S^{}_0 S^{}_3 U^\prime_0 D^{}_N U^{\prime T}_0 A^T_{3}\right)^{-1}
\left( R^{}_{1} D^{}_N U^{\prime T}_0 S^T_{3} S^T_{0} +
R^{}_{2} D^{}_S B^T_{3} S^T_0 \right)^T
\nonumber \\
\hspace{-0.2cm} & \simeq & \hspace{-0.2cm}
M^{}_{\rm D} \left(M^T_S\right)^{-1} \varepsilon^T \; ,
\nonumber \\
R^{}_{1} D^{}_N R^T_{2} + R^{}_{2} D^{}_S R^T_{2}
\hspace{-0.2cm} & \simeq & \hspace{-0.2cm}
R^{}_{1} D^{}_N \left( R^{}_2 B^{-1}_3 S^{}_3 U^{\prime}_0\right)^T + R^{}_{2} D^{}_S R^T_{2}
\nonumber \\
\hspace{-0.2cm} & \simeq & \hspace{-0.2cm}
\left( R^{}_{1} D^{}_N U^{\prime T}_0 S^T_{3} S^T_{0} + R^{}_{2} D^{}_S B^T_{3} S^T_0 \right) \left( 2R^{}_{3} D^{}_S B^T_{3} S^T_0 \right)^{-1}
\left( 2R^{}_{2} D^{}_S R^T_{3} \right)^T
\nonumber \\
\hspace{-0.2cm} & \simeq & \hspace{-0.2cm}
\varepsilon \left(M^{}_S\right)^{-1} M^T_{\rm D} \; .
\end{eqnarray}
Combining Eq.~(36) with Eq.~(34) allows us to obtain the linear seesaw formula
\begin{eqnarray}
M^{}_\nu \hspace{-0.2cm} & \simeq & \hspace{-0.2cm}
- R^{}_{1} D^{}_N R^T_{1} - R^{}_{2} D^{}_S R^T_{2}
\nonumber \\
\hspace{-0.2cm} & \simeq & \hspace{-0.2cm}
- M^{}_{\rm D} \left( M^{T}_S \right)^{-1} \varepsilon^T_{} -
\varepsilon^{}_{} \left( M^{}_S \right)^{-1} M^{T}_{\rm D} \; . \hspace{0.8cm}
\end{eqnarray}
This result is fully consistent with Eq.~(4).

\section{Jarlskog invariants and unitarity nonagons}
\label{section:4}

\subsection{The Jarlskog invariants}

In the standard three-flavor scheme without any new degrees of freedom, the strength of
CP violation in neutrino oscillations is measured by the unique Jarlskog invariant
\cite{Jarlskog:1985ht}
\begin{eqnarray}
{\cal J}^{}_0 = {\cal J}^{ij}_{\alpha \beta} \equiv {\rm Im}
\left( U^{}_{\alpha i} U^{}_{\beta j} U^{*}_{\alpha j} U^{*}_{\beta i}\right) \; ,
\end{eqnarray}
where the Latin (or Greek) indices $i$ and $j$ (or $\alpha$ and $\beta$) run
cyclically over $1$, $2$ and $3$ (or $e$, $\mu$ and $\tau$). Since $U = U^{}_0$ holds
in this case, the parametrization of $U^{}_0$ in Eq.~(9) leads us to
\begin{eqnarray}
{\cal J}^{}_0 = c^{}_{12} s^{}_{12} c^{2}_{13} s^{}_{13} c^{}_{23} s^{}_{23}
\sin \delta \; ,
\end{eqnarray}
where $\delta = \delta^{}_{13} - \delta^{}_{12} - \delta^{}_{23}$ is sometimes
referred to as the ``Dirac" CP-violating phase. Note that this phase parameter
may also manifest itself in the lepton-number-violating neutrino-antineutrino
oscillations \cite{Xing:2013ty,Xing:2013woa}, simply because it is a nontrivial
CP phase of Majorana neutrinos.

In the inverse or linear seesaw scenario, however, the $3\times 3$ PMNS matrix
$U = A^{}_{2} A^{}_{1} U^{}_0$ is not exactly unitary. As a consequence, some
additional effects of CP violation described by the deviations of
${\cal J}^{ij}_{\alpha \beta}$ from ${\cal J}^{}_0$
will manifest themselves in neutrino oscillations. To explicitly
calculate the nine Jarlskog invariants ${\cal J}^{ij}_{\alpha \beta}$, let
us rewrite the approximate expressions of $A^{}_1$ in Eq.~(B1) and $A^{}_2$
in Eq.~(B3) as follows:
\begin{eqnarray}
A^{}_{1\left(2\right)} \simeq
\left( \begin{matrix} 1 & 0 & 0 \cr -X^{}_{1 \left( 2 \right) } & 1 & 0 \cr
-Y^{}_{1 \left( 2 \right) } & -Z^{}_{1 \left( 2 \right) } & 1 \end{matrix} \right) \; ,
\end{eqnarray}
where only the leading term of each matrix element is kept, and
\begin{eqnarray}
X^{}_{1} \hspace{-0.2cm} & \equiv & \hspace{-0.2cm}
\hat{s}^{}_{14} \hat{s}^*_{24} + \hat{s}^{}_{15} \hat{s}^*_{25}
+ \hat{s}^{}_{16} \hat{s}^*_{26} \; ,
\nonumber \\
Y^{}_{1} \hspace{-0.2cm} & \equiv & \hspace{-0.2cm}
\hat{s}^{}_{14} \hat{s}^*_{34} + \hat{s}^{}_{15} \hat{s}^*_{35}
+ \hat{s}^{}_{16} \hat{s}^*_{36} \; ,
\nonumber \\
Z^{}_{1} \hspace{-0.2cm} & \equiv & \hspace{-0.2cm}
\hat{s}^{}_{24} \hat{s}^*_{34} + \hat{s}^{}_{25} \hat{s}^*_{35}
+ \hat{s}^{}_{26} \hat{s}^*_{36} \; , \hspace{0.8cm}
\end{eqnarray}
together with
\begin{eqnarray}
X^{}_{2} \hspace{-0.2cm} & \equiv & \hspace{-0.2cm}
\hat{s}^{}_{17} \hat{s}^*_{27} + \hat{s}^{}_{18} \hat{s}^*_{28}
+ \hat{s}^{}_{19} \hat{s}^*_{29} \; ,
\nonumber \\
Y^{}_{2} \hspace{-0.2cm} & \equiv & \hspace{-0.2cm}
\hat{s}^{}_{17} \hat{s}^*_{37} + \hat{s}^{}_{18} \hat{s}^*_{38}
+ \hat{s}^{}_{19} \hat{s}^*_{39} \; ,
\nonumber \\
Z^{}_{2} \hspace{-0.2cm} & \equiv & \hspace{-0.2cm}
\hat{s}^{}_{27} \hat{s}^*_{37} + \hat{s}^{}_{28} \hat{s}^*_{38}
+ \hat{s}^{}_{29} \hat{s}^*_{39} \; .  \hspace{0.8cm}
\end{eqnarray}
Now that $A^{}_1$ and $A^{}_2$ signify the departure of $U$ from $U^{}_0$,
the magnitudes of $X^{}_{1\left(2\right)}$, $Y^{}_{1\left(2\right)}$ and
$Z^{}_{1\left(2\right)}$ are at most of the percent level. On the other hand,
the smallest flavor mixing angle of $U^{}_0$ is $\theta^{}_{13}$, and its
size is about $0.16$. So it is reasonable to omit the terms of
${\cal O} (X^{}_{1(2)}s^{}_{13})$, ${\cal O} (Y^{}_{1(2)}s^{}_{13})$,
${\cal O} (Z^{}_{1(2)}s^{}_{13})$, ${\cal O} (X^{2}_{1(2)})$,
${\cal O} (Y^{2}_{1(2)})$ and ${\cal O}(Z^{2}_{1(2)})$ in calculating
${\cal J}^{ij}_{\alpha\beta}$. Our results are
\begin{eqnarray}
{\cal J}^{12}_{e \mu} \hspace{-0.2cm} & \simeq & \hspace{-0.2cm}
{\cal J}^{}_0 + c^{}_{12} s^{}_{12} c^{}_{23} {\rm Im} \Big[\big( X^{}_{1} + X^{}_{2}
\big) e^{-{\rm i} \delta^{}_{12}}\Big] \; ,
\nonumber \\
{\cal J}^{12}_{\tau e} \hspace{-0.2cm} & \simeq & \hspace{-0.2cm}
{\cal J}^{}_0 + c^{}_{12} s^{}_{12} s^{}_{23} {\rm Im} \Big[\big( Y^{}_{1} + Y^{}_{2}
\big) e^{-{\rm i} \left( \delta^{}_{12} + \delta^{}_{23} \right)}\Big] \; ,
\nonumber \\
{\cal J}^{12}_{\mu \tau} \hspace{-0.2cm} & \simeq & \hspace{-0.2cm}
{\cal J}^{}_0 + c^{}_{12} s^{}_{12} c^{}_{23} s^{}_{23} \Big\{ s^{}_{23} {\rm Im}
\Big[\big( X^{}_{1} + X^{}_{2} \big) e^{-{\rm i} \delta^{}_{12}}\Big]
+ c^{}_{23} {\rm Im} \Big[\big( Y^{}_{1} + Y^{}_{2} \big) e^{-{\rm i}
\left( \delta^{}_{12} + \delta^{}_{23} \right)} \Big]\Big\} \; ,
\nonumber \\
{\cal J}^{23}_{\mu \tau} \hspace{-0.2cm} & \simeq & \hspace{-0.2cm}
{\cal J}^{}_0 + c^{}_{12} c^{}_{23} s^{}_{23} \Big\{s^{}_{12} s^{}_{23}
{\rm Im} \Big[ \big(X^{}_{1} + X^{}_{2} \big) e^{-{\rm i} \delta^{}_{12}}\Big]
+ s^{}_{12} c^{}_{23} {\rm Im} \Big[\big( Y^{}_{1} + Y^{}_{2} \big)
e^{-{\rm i} \left( \delta^{}_{12} + \delta^{}_{23} \right) }\Big]
\nonumber \\
\hspace{-0.2cm} & & \hspace{-0.2cm}
+ c^{}_{12} {\rm Im} \Big[\big( Z^{}_{1} + Z^{}_{2} \big)
e^{-{\rm i} \delta^{}_{23}} \Big]\Big\} \; ,
\nonumber \\
{\cal J}^{31}_{\mu \tau} \hspace{-0.2cm} & \simeq & \hspace{-0.2cm}
{\cal J}^{}_0 + s^{}_{12} c^{}_{23} s^{}_{23} \Big\{c^{}_{12} s^{}_{23}
{\rm Im} \Big[\big( X^{}_{1} + X^{}_{2} \big) e^{-{\rm i} \delta^{}_{12}}\Big]
+ c^{}_{12} c^{}_{23} {\rm Im} \Big[\big( Y^{}_{1} + Y^{}_{2} \big)
e^{-{\rm i} \left( \delta^{}_{12} + \delta^{}_{23} \right)}\Big] \hspace{0.5cm}
\nonumber \\
\hspace{-0.2cm} & & \hspace{-0.2cm}
- s^{}_{12} {\rm Im} \Big[\big( Z^{}_{1} + Z^{}_{2} \big)
e^{-{\rm i} \delta^{}_{23}}\Big]\Big\} \; ,
\end{eqnarray}
and ${\cal J}^{23}_{e \mu} \simeq {\cal J}^{31}_{e \mu} \simeq {\cal J}^{23}_{\tau e}
\simeq {\cal J}^{31}_{\tau e} \simeq {\cal J}^{}_{0}$. One can see that five of the
nine Jarlskog invariants are sensitive to the active-sterile flavor mixing angles
and the associated CP-violating phases.

As for the probabilities of neutrino oscillations in vacuum, let us assume that
all the sterile particles are heavy enough and hence kinematically forbidden to
participate in a realistic long-baseline oscillation process. In this case one
does not have to worry about the masses of those hypothetical particles
and their differences from the masses of three active neutrinos, and the Jarlskog
invariants calculated above determine the CP-violating asymmetry
${\cal A}^{}_{\alpha\beta} \equiv P(\nu^{}_\alpha \to \nu^{}_\beta) -
P(\overline\nu^{}_\alpha \to \overline\nu^{}_\beta)$
(for $\alpha, \beta = e, \mu, \tau$) as follows \cite{Xing:2011ur}:
\begin{eqnarray}
{\cal A}^{}_{\alpha\beta}
\hspace{-0.2cm} & = & \hspace{-0.2cm}
-\frac{4}{\left(UU^\dagger\right)^{}_{\alpha\alpha}
\left(UU^\dagger\right)^{}_{\beta\beta}}
\sum^{}_{i<j} {\cal	J}^{ij}_{\alpha\beta} \sin \frac{\Delta m^{2}_{ij}L}{2E}
\nonumber \\
\hspace{-0.2cm} & \simeq & \hspace{-0.2cm}
4 \left[{\cal J}^{12}_{\alpha\beta} \sin\frac{\Delta m^{2}_{21}L}{2E} -
{\cal J}^{31}_{\alpha\beta} \sin\frac{\Delta m^{2}_{31}L}{2E}
+ {\cal J}^{23}_{\alpha\beta} \sin\frac{\Delta m^{2}_{32}L}{2E} \right] \; ,
\hspace{0.5cm}
\end{eqnarray}
where $\Delta m^2_{ij} \equiv m^2_i - m^2_j$ (for $i,j = 1,2,3$) are defined, $E$
and $L$ stand respectively for the neutrino beam energy and baseline length, and the
approximations $A^{}_{1} A^\dagger_{1} \simeq I$ and $A^{}_{2} A^\dagger_{2} \simeq I$
have been used. In practice, however, terrestrial matter effects must be taken into
account for a long-baseline neutrino oscillation experiment which is sensitive to
leptonic CP violation (see, e.g., Refs.~\cite{Fernandez-Martinez:2007iaa,
Goswami:2008mi,Luo:2008vp,Antusch:2009pm,Li:2015oal,Ohlsson:2012kf,Ohlsson:2013ip}).
That is why a careful analysis of the non-unitary flavor mixing effects and matter
effects is needed for a realistic long-baseline neutrino (or antineutrino) oscillation
experiment \cite{Li:2018jgd}, no matter whether CP violation is concerned or not.

\subsection{The unitarity nonagons}

It is well known that six orthogonality relations of the $3\times 3$ unitary PMNS
matrix $U = U^{}_0$ define six triangles in the complex plane, which are referred
to as the unitarity triangles \cite{Fritzsch:1999ee,Aguilar-Saavedra:2000jom}.
Their areas are all equal to ${\cal J}^{}_0/2$, thanks to the unitarity of $U^{}_0$.
Three of the six triangles, defined by
\begin{eqnarray}	
\triangle^{}_\tau & : & U^{}_{e 1} U^*_{\mu 1} + U^{}_{e 2}
	U^*_{\mu 2} + U^{}_{e 3} U^*_{\mu 3} = 0 \;,
\nonumber \\
\triangle^{}_e & : & U^{}_{\mu 1} U^*_{\tau 1} + U^{}_{\mu 2}
	U^*_{\tau 2} + U^{}_{\mu 3} U^*_{\tau 3} = 0 \;, \hspace{0.5cm}
\nonumber \\
\triangle^{}_\mu & : & U^{}_{\tau 1} U^*_{e 1} + U^{}_{\tau 2}
	U^*_{e 2} + U^{}_{\tau 3} U^*_{e 3} = 0 \; ,
\end{eqnarray}
where $U = U^{}_0$ is implied,
are directly associated with leptonic CP violation in neutrino oscillations.

Given three sterile neutrinos and three extra neutral fermions which slightly mix with
the active neutrinos, the corresponding $9\times 9$ unitary flavor mixing
matrix $\cal U$ allows us to totally define 36 nonagons in the complex plane.
Among them, only three unitarity nonagons defined by the orthogonality conditions
$\left( {\cal UU}^{\dagger} \right)^{}_{e\mu} =0$,
$\left( {\cal UU}^{\dagger} \right)^{}_{\mu\tau} =0$ and
$\left( {\cal UU}^{\dagger} \right)^{}_{\tau e} =0$ are relevant to the flavor
oscillations of three active neutrinos. Since the sides associated with sterile
flavors must be very short as compared with the sides $|U^{}_{\alpha i}U^*_{\beta i}|$
for $(\alpha, \beta) = (e, \mu)$, $(\mu, \tau)$, $(\tau, e)$ and $i=1,2,3$,
these three unitarity nonagons can be regarded as the deformed versions of
unitarity triangles $\triangle^{}_\tau$, $\triangle^{}_e$ and $\triangle^{}_\mu$.
Taking account of Eqs.~(40)---(42) in this case, the three unitarity nonagons
under discussion can therefore be expressed as
\begin{eqnarray}	
\triangle^{\prime}_\tau & : & U^{}_{e 1} U^*_{\mu 1} + U^{}_{e 2}
	U^*_{\mu 2} + U^{}_{e 3} U^*_{\mu 3} \simeq -X^*_{1}-X^*_{2} \;, \hspace{0.5cm}
\nonumber \\
\triangle^{\prime}_e & : & U^{}_{\mu 1} U^*_{\tau 1} + U^{}_{\mu 2}
	U^*_{\tau 2} + U^{}_{\mu 3} U^*_{\tau 3} \simeq -Z^*_{1}-Z^*_{2} \;,
\nonumber \\
\triangle^{\prime}_\mu & : & U^{}_{\tau 1} U^*_{e 1} + U^{}_{\tau 2}
	U^*_{e 2} + U^{}_{\tau 3} U^*_{e 3} \simeq -Y^{}_{1}-Y^{}_{2} \; .
\end{eqnarray}
Comparing Eq.~(46) with Eq.~(45), one can see that the new sides originating from those
sterile degrees of freedom are all of ${\cal O}(s^2_{ij})$ with $i = 1,2,3$ and
$j = 4,5,6,7,8,9$. That is why we use $\triangle^\prime_\alpha$ to denote the
unitarity nonagon defined by $\left( {\cal UU}^{\dagger} \right)^{}_{\beta\gamma} =0$
(for $\alpha$, $\beta$ and $\gamma$ running cyclically over $e$, $\mu$ and $\tau$),
simply because it deviates only slightly from $\triangle^{}_\alpha$.

\begin{figure}[htbp]
{\hspace{-2.3cm}\includegraphics[width=21cm]{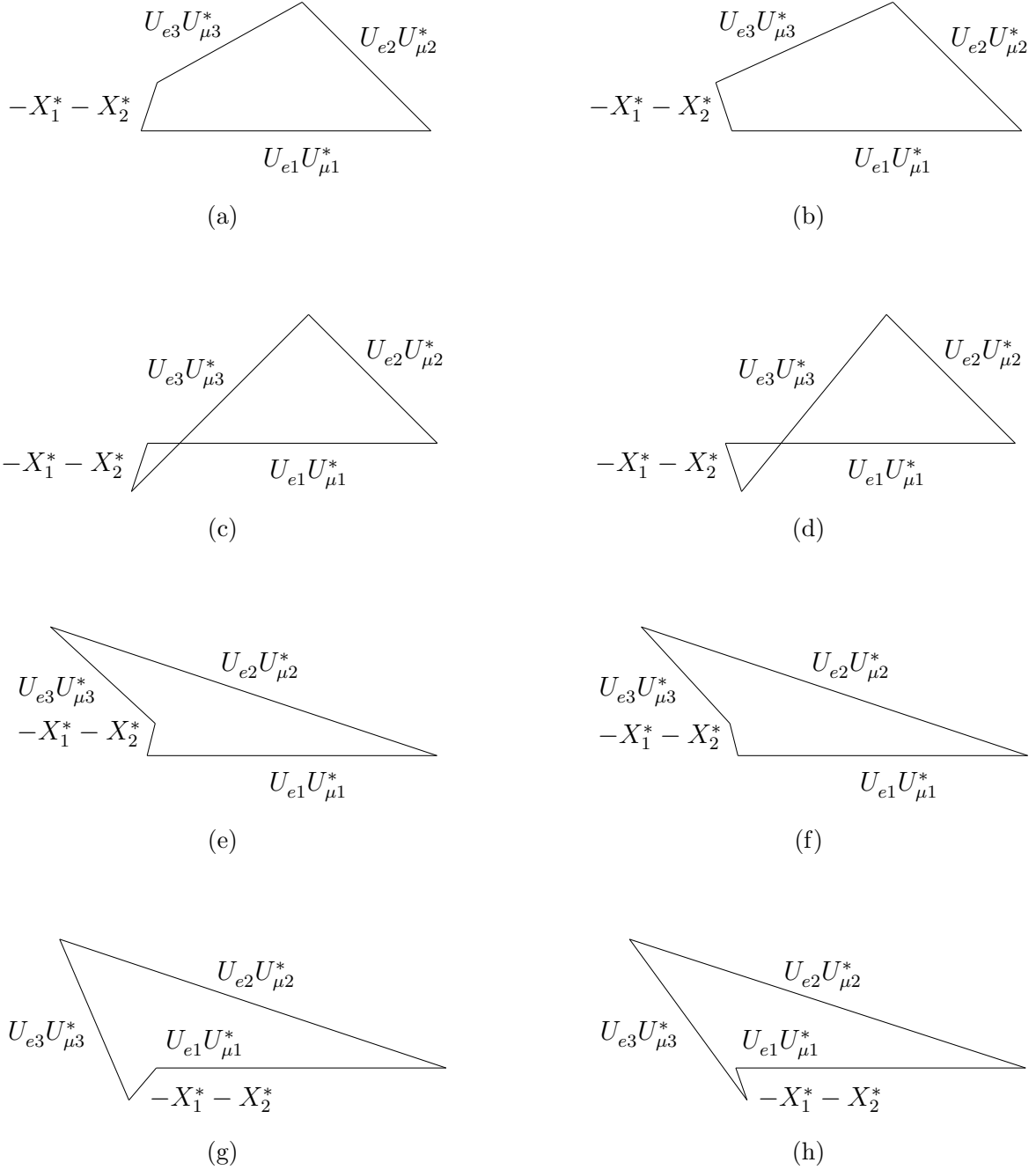}}
\vspace{-8.7cm}
\caption{A schematic illustration of possible shapes of the effective
unitarity quadrangle $\triangle^{\prime}_\tau$ as an example in the complex plane, where
the three long sides correspond to $U^{}_{e 1} U^*_{\mu 1}$, $U^{}_{e 2}U^*_{\mu 2}$
and $U^{}_{e 3} U^*_{\mu 3}$, and the fourth (effective) side is measured by
$-X^*_{1}-X^*_{2}$.}
\label{FIG1}
\end{figure}

In the lack of definite information about the active-sterile flavor mixing parameters
(i.e., $\theta^{}_{ij}$ and $\delta^{}_{ij}$ for $i = 1,2,3$ and
$j = 4,5,6,7,8,9$), one may simply treat the new sides of $\triangle^\prime_\alpha$
(for $\alpha = e$, $\mu$ or $\tau$) on the right-hand side of Eq.~(46)
as an effective ``single" side. In this case the unitarity nonagon is reduced to
an effective unitarity quadrangle whose shortest side signifies the existence of
sterile neutrinos and extra neutral fermions, as schematically illustrated
by Fig.~\ref{FIG1} for $\triangle^\prime_\tau$.
For simplicity, we have made the choice that the side $U^{}_{e 1} U^*_{\mu 1}$ lies
in the horizontal direction and forms a sharp angle to the side
$U^{}_{e 2} U^*_{\mu 2}$ in Fig.~\ref{FIG1}. Then the side $U^{}_{e 3} U^*_{\mu 3}$
may form either a sharp angle or an obtuse angle to the horizontal side, and the
shortest side defined by $-X^*_{1}-X^*_{2}$ is likely to link these two longer
sides in several different topologies.

To constrain the shortest side of each effective unitarity quadrangle, one may study
those lepton-flavor-violating processes such as neutrino oscillations and radiative
decays of charged leptons. In the latter case the new degrees of freedom can mediate
the one-loop $\alpha^- \to \beta^- + \gamma$ transitions for $(\alpha, \beta)
= (\mu, e)$, $(\tau, e)$ and $(\tau, \mu)$, and the relevant loop functions depend on
the masses of three sterile neutrinos and three extra neutral fermions (i.e.,
$M^{}_i$ and $M^\prime_i$ for $i=1,2,3$) \cite{Camara:2020efq,Xing:2020ivm,Zhang:2021tsq}.
A systematic analysis of such lepton-flavor-violating processes in the inverse and
linear seesaw scenarios will be done elsewhere.

\section{Concluding remarks}

The inverse and linear seesaw scenarios are two simple extensions of the canonical
(type-I) seesaw mechanism aiming to lower the mass scales of those hypothetical particles
and hence enhance the experimental testability. In this regard a high price that one
has to pay is the introduction of three species of extra neutral fermions besides
three species of sterile neutrinos. How to determine or constrain the flavor structure
of such a complicated seesaw scenario and describe the corresponding flavor mixing
pattern turns out to be a highly nontrivial issue.

Instead of trying to reduce the number of free parameters by imposing some kind of
flavor symmetry or empirical assumptions on the texture of the $9\times 9$ mass matrix
(or its $3\times 3$ submatrices) in either the inverse seesaw scenario or the linear
seesaw scenario \cite{Xing:2019vks,King:2003jb,Altarelli:2010gt,Feruglio:2019ybq},
here we have focused on a full description of the $9\times 9$ active-sterile flavor
mixing matrix in terms of 36 rotation angles and 36 CP-violating phases. Such a
generic and model-independent work is certainly new. The most salient feature of our
parametrization that the primary unitary $3\times 3$ flavor mixing submatrices of
three active neutrinos, three sterile neutrinos and three extra neutral fermions,
which would look like three isolated islands if the Yukawa-like interactions
among them were absent, are linked and modified by the intermediate flavor mixing
matrices. This approach proves to be useful for describing possible deviations of
the $3\times 3$ PMNS matrix of three active neutrinos from its unitary limit, as
it allows us to calculate the effective Jarlskog invariants and the deformed
unitarity triangles in a good approximation and without loss of any generality.

Starting from our generic result, one may easily arrive at a full Euler-like
parametrization of the $7\times 7$ active-sterile flavor mixing matrix in the
so-called {\it minimal} inverse seesaw model which contains only two species
of sterile neutrinos and two species of extra neutral fermions (see, e.g.,
Refs.~\cite{Malinsky:2009df,Camara:2020efq,Hirsch:2009ra,Mondal:2012jv,Abada:2014vea,
Abada:2014zra,CarcamoHernandez:2019eme}). A similar simplification can also be applied
to the so-called {\it minimal} linear seesaw model \cite{Matute:2021osr} and
the study of its various phenomenological consequences.

Finally, it is also worth pointing out that our exact Euler-like parametrization of
the $n\times n$ unitary matrix (for $3 \leq n \leq 9$) is likely to find applications
in some other physical systems.

\section*{Acknowledgements}

We are indebted to Jian-Wei Mei and Shun Zhou for useful discussions.
This work is supported in part by the National Natural
Science Foundation of China under grant No. 12075254, grant
No. 11775231 and grant No. 11835013.

\allowdisplaybreaks[4]
\appendix

\setcounter{equation}{0}
\renewcommand\theequation{A.\arabic{equation}}
	
\section{The unitary conditions of $\cal U$}	
\label{appendix A}

Given the expression of the $9\times 9$ active-sterile flavor mixing matrix
$\cal U$ in Eq.~(16), one may obtain either the six unitary conditions for
its $3\times 3$ submatrices from ${\cal U} {\cal U}^\dagger = I^{}_{9\times 9}$:
\begin{eqnarray}
(1) \hspace{-0.2cm} && \hspace{-0.2cm}
A^{}_2 A^{}_1 A^{\dagger}_1 A^{\dagger}_2 + A^{}_2 R^{}_1 R^{\dagger}_1
A^{\dagger}_2 + R^{}_2 R^{\dagger}_2 = I \; ,
\nonumber \\
(2) \hspace{-0.2cm} && \hspace{-0.2cm}
\big( R^{}_3 S^{}_2 A^{}_1 + A^{}_3 U^{\prime}_0 S^{}_1 \big)
\big( A^{\dagger}_1 S^{\dagger}_2 R^{\dagger}_3 + S^{\dagger}_1
U^{\prime\dagger}_0 A^{\dagger}_3 \big)
+ \big( R^{}_3 S^{}_2 R^{}_1 + A^{}_3 U^{\prime}_0 B^{}_1 \big)
\big( R^{\dagger}_1 S^{\dagger}_2 R^{\dagger}_3 + B^{\dagger}_1
U^{\prime\dagger}_0 A^{\dagger}_3 \big)
\nonumber \\
\hspace{-0.2cm} && \hspace{-0.2cm}
+ R^{}_3 B^{}_2 B^{\dagger}_2 R^{\dagger}_3 = I \; ,
\nonumber \\
(3) \hspace{-0.2cm} && \hspace{-0.2cm}
\big( B^{}_3 S^{}_2 A^{}_1 + S^{}_3 U^{\prime}_0 S^{}_1 \big)
\big( A^{\dagger}_1 S^{\dagger}_2 B^{\dagger}_3 + S^{\dagger}_1
U^{\prime\dagger}_0 S^{\dagger}_3 \big)
+ \big( B^{}_3 S^{}_2 R^{}_1 + S^{}_3 U^{\prime}_0 B^{}_1 \big)
\big( B^{\dagger}_1 U^{\prime\dagger}_0 S^{\dagger}_3 +
R^{\dagger}_1 S^{\dagger}_2 B^{\dagger}_3 \big)
\nonumber \\
\hspace{-0.2cm} && \hspace{-0.2cm}
+ B^{}_3 B^{}_2 B^{\dagger}_2 B^{\dagger}_3 = I \; ,
\nonumber \\
(4) \hspace{-0.2cm} && \hspace{-0.2cm}
A^{}_2 A^{}_1 \big( A^{\dagger}_1 S^{\dagger}_2 R^{\dagger}_3 +
S^{\dagger}_1 U^{\prime\dagger}_0 A^{\dagger}_3 \big)
+ A^{}_2 R^{}_1 \big( R^{\dagger}_1 S^{\dagger}_2 R^{\dagger}_3 +
B^{\dagger}_1 U^{\prime\dagger}_0 A^{\dagger}_3 \big)
+ R^{}_2 B^{\dagger}_2 R^{\dagger}_3 = 0 \; ,
\nonumber \\
(5) \hspace{-0.2cm} && \hspace{-0.2cm}
\big( R^{}_3 S^{}_2 A^{}_1 + A^{}_3 U^{\prime}_0 S^{}_1 \big)
\big( A^{\dagger}_1 S^{\dagger}_2 B^{\dagger}_3 + S^{\dagger}_1
U^{\prime\dagger}_0 S^{\dagger}_3 \big)
+ \big( R^{}_3 S^{}_2 R^{}_1 + A^{}_3 U^{\prime}_0 B^{}_1 \big)
\big( B^{\dagger}_1 U^{\prime\dagger}_0 S^{\dagger}_3 +
R^{\dagger}_1 S^{\dagger}_2 B^{\dagger}_3 \big)
\nonumber \\
\hspace{-0.2cm} && \hspace{-0.2cm}
+ R^{}_3 B^{}_2 B^{\dagger}_2 B^{\dagger}_3 = 0 \;,
\nonumber \\
(6) \hspace{-0.2cm} && \hspace{-0.2cm}
\big( B^{}_3 S^{}_2 A^{}_1 + S^{}_3 U^{\prime}_0 S^{}_1 \big)
A^{\dagger}_1 A^{\dagger}_2 + \big( B^{}_3 S^{}_2 R^{}_1 + S^{}_3
U^{\prime}_0 B^{}_1 \big) R^{\dagger}_1 A^{\dagger}_2
+ B^{}_3 B^{}_2 R^{\dagger}_2 = 0 \; ;
\end{eqnarray}
or the six unitary conditions for its $3\times 3$ submatrices from
${\cal U}^\dagger {\cal U} = I^{}_{9\times 9}$:
\begin{eqnarray}
(1) \hspace{-0.2cm} && \hspace{-0.2cm}
U^{\dagger}_0 A^{\dagger}_1 A^{\dagger}_2 A^{}_2 A^{}_1 U^{}_0 +
U^{\dagger}_0 \big( A^{\dagger}_1 S^{\dagger}_2 R^{\dagger}_3 +
S^{\dagger}_1 U^{\prime\dagger}_0 A^{\dagger}_3 \big)
\big( R^{}_3 S^{}_2 A^{}_1 + A^{}_3 U^{\prime}_0 S^{}_1 \big) U^{}_0
\nonumber \\
\hspace{-0.2cm} && \hspace{-0.2cm}
+ U^{\dagger}_0 \big( A^{\dagger}_1 S^{\dagger}_2 B^{\dagger}_3 +
S^{\dagger}_1 U^{\prime\dagger}_0 S^{\dagger}_3 \big)
\big( B^{}_3 S^{}_2 A^{}_1 + S^{}_3 U^{\prime}_0 S^{}_1 \big) U^{}_0 = I \; ,
\nonumber \\
(2) \hspace{-0.2cm} && \hspace{-0.2cm}
R^{\dagger}_1 A^{\dagger}_2 A^{}_2 R^{}_1 + \big( R^{\dagger}_1 S^{\dagger}_2
R^{\dagger}_3 + B^{\dagger}_1 U^{\prime\dagger}_0 A^{\dagger}_3 \big)
\big( R^{}_3 S^{}_2 R^{}_1 + A^{}_3 U^{\prime}_0 B^{}_1 \big)
\nonumber \\
\hspace{-0.2cm} && \hspace{-0.2cm}
+ \big( B^{\dagger}_1 U^{\prime\dagger}_0 S^{\dagger}_3 +
R^{\dagger}_1 S^{\dagger}_2 B^{\dagger}_3 \big)
\big( B^{}_3 S^{}_2 R^{}_1 + S^{}_3 U^{\prime}_0 B^{}_1 \big) = I \; ,
\nonumber \\
(3) \hspace{-0.2cm} && \hspace{-0.2cm}
R^{\dagger}_2 R^{}_2 + B^{\dagger}_2 R^{\dagger}_3 R^{}_3 B^{}_2 +
B^{\dagger}_2 B^{\dagger}_3 B^{}_3 B^{}_2 = I \; ,
\nonumber \\
(4) \hspace{-0.2cm} && \hspace{-0.2cm}
A^{\dagger}_1 A^{\dagger}_2 A^{}_2 R^{}_1 +
\big( A^{\dagger}_1 S^{\dagger}_2 R^{\dagger}_3 + S^{\dagger}_1
U^{\prime\dagger}_0 A^{\dagger}_3 \big)
\big( R^{}_3 S^{}_2 R^{}_1 + A^{}_3 U^{\prime}_0 B^{}_1 \big)
\nonumber \\
\hspace{-0.2cm} && \hspace{-0.2cm}
+ \big( A^{\dagger}_1 S^{\dagger}_2 B^{\dagger}_3 + S^{\dagger}_1
U^{\prime\dagger}_0 S^{\dagger}_3 \big) \big( B^{}_3 S^{}_2 R^{}_1
+ S^{}_3 U^{\prime}_0 B^{}_1 \big) = 0 \; ,
\nonumber \\
(5) \hspace{-0.2cm} && \hspace{-0.2cm}
R^{\dagger}_1 A^{\dagger}_2 R^{}_2 + \big( R^{\dagger}_1 S^{\dagger}_2
R^{\dagger}_3 + B^{\dagger}_1 U^{\prime\dagger}_0 A^{\dagger}_3 \big)
R^{}_3 B^{}_2
+ \big( B^{\dagger}_1 U^{\prime\dagger}_0 S^{\dagger}_3 +
R^{\dagger}_1 S^{\dagger}_2 B^{\dagger}_3 \big) B^{}_3 B^{}_2 = 0 \; ,
\hspace{2.3cm}
\nonumber \\
(6) \hspace{-0.2cm} && \hspace{-0.2cm}
R^{\dagger}_2 A^{}_2 A^{}_1 + B^{\dagger}_2 R^{\dagger}_3 \big( R^{}_3
S^{}_2 A^{}_1 + A^{}_3 U^{\prime}_0 S^{}_1 \big)
+ B^{\dagger}_2 B^{\dagger}_3 \big( B^{}_3 S^{}_2 A^{}_1 +
S^{}_3 U^{\prime}_0 S^{}_1 \big) = 0 \; .
\end{eqnarray}
Switching off the extra neutral fermion sector, for example, we arrive
at $A^{}_2 = I$ and $R^{}_2 = 0$. In this case the first relation in
Eq.~(A.1) can be simplified to $A^{}_1 A^\dagger_1 + R^{}_1 R^\dagger_1 = I$,
which is valid for the canonical seesaw mechanism
\cite{Xing:2011ur,Xing:2007zj,Xing:2019vks}.

\setcounter{equation}{0}
\renewcommand\theequation{B.\arabic{equation}}

\section{The approximate forms of $A^{}_{1,2}$, $B^{}_{1,2}$, $R^{}_{1,2}$
and $S^{}_{1,2}$}
\label{appendix B}

Given the very fact that the eighteen active-sterile flavor mixing angles
$\theta^{}_{1j}$, $\theta^{}_{2j}$ and $\theta^{}_{3j}$ (for $j=4, 5, \cdots, 9$)
must be strongly suppressed in magnitude, one may simplify the four
$3\times 3$ matrices $A^{}_{1}$, $B^{}_{1}$, $R^{}_{1}$ and $S^{}_{1}$
in Eqs.~(10) and (11) to the following forms:
\begin{eqnarray}
A^{}_1 \hspace{-0.2cm} & \simeq & \hspace{-0.2cm}
I - \left( \begin{matrix} \frac{1}{2} \left( s^2_{14} +
	s^2_{15} + s^2_{16} \right) & 0 & 0 \cr \hat{s}^{}_{14}
	\hat{s}^*_{24} + \hat{s}^{}_{15} \hat{s}^*_{25} + \hat{s}^{}_{16}
	\hat{s}^*_{26} & \frac{1}{2} \left( s^2_{24} + s^2_{25} + s^2_{26}
	\right) & 0 \cr \hat{s}^{}_{14} \hat{s}^*_{34} + \hat{s}^{}_{15}
	\hat{s}^*_{35} + \hat{s}^{}_{16} \hat{s}^*_{36} & \hat{s}^{}_{24}
	\hat{s}^*_{34} + \hat{s}^{}_{25} \hat{s}^*_{35} + \hat{s}^{}_{26}
	\hat{s}^*_{36} & \frac{1}{2} \left( s^2_{34} +
	s^2_{35} + s^2_{36} \right) \cr \end{matrix} \right) \;,
\nonumber \\
B^{}_1 \hspace{-0.2cm} & \simeq & \hspace{-0.2cm}
I - \left( \begin{matrix} \frac{1}{2} \left( s^2_{14} +
	s^2_{24} + s^2_{34} \right) & 0 & 0 \cr \hat{s}^{*}_{14}
	\hat{s}^{}_{15} + \hat{s}^{*}_{24} \hat{s}^{}_{25} + \hat{s}^{*}_{34}
	\hat{s}^{}_{35} & \frac{1}{2} \left( s^2_{15} + s^2_{25} + s^2_{35}
	\right) & 0 \cr \hat{s}^{*}_{14} \hat{s}^{}_{16} + \hat{s}^{*}_{24}
	\hat{s}^{}_{26} + \hat{s}^{*}_{34} \hat{s}^{}_{36} & \hat{s}^{*}_{15}
	\hat{s}^{}_{16} + \hat{s}^{*}_{25} \hat{s}^{}_{26} + \hat{s}^{*}_{35}
	\hat{s}^{}_{36} & \frac{1}{2} \left( s^2_{16} +
	s^2_{26} + s^2_{36} \right) \cr \end{matrix} \right) \;, \hspace{0.5cm}
\end{eqnarray}
where the terms of ${\cal O}(s^4_{ij})$ (for $i=1,2,3$ and $j=4,5,6$)
have been omitted; and
\begin{eqnarray}
S^{}_1 \simeq -R^\dagger_1 \simeq - \left( \begin{matrix} \hat{s}^{}_{14} & \hat{s}^{}_{24}
	& \hat{s}^{}_{34} \cr \hat{s}^{}_{15} & \hat{s}^{}_{25} &
	\hat{s}^{}_{35} \cr \hat{s}^{}_{16} & \hat{s}^{}_{26} &
	\hat{s}^{}_{36} \cr \end{matrix} \right) \;,
\end{eqnarray}
where the terms of ${\cal O}(s^3_{ij})$ (for $i=1,2,3$ and $j=4,5,6$)
have been omitted. Similarly, the four $3\times 3$ matrices $A^{}_{2}$,
$B^{}_{2}$, $R^{}_{2}$ and $S^{}_{2}$ in Eqs.~(12) and (13) can be simplified as follows:
\begin{eqnarray}
A^{}_2 \hspace{-0.2cm} & \simeq & \hspace{-0.2cm}
I - \left( \begin{matrix} \frac{1}{2} \left( s^2_{17} +
	s^2_{18} + s^2_{19} \right) & 0 & 0 \cr \hat{s}^{}_{17}
	\hat{s}^*_{27} + \hat{s}^{}_{18} \hat{s}^*_{28} + \hat{s}^{}_{19}
	\hat{s}^*_{29} & \frac{1}{2} \left( s^2_{27} + s^2_{28} + s^2_{29}
	\right) & 0 \cr \hat{s}^{}_{17} \hat{s}^*_{37} + \hat{s}^{}_{18}
	\hat{s}^*_{38} + \hat{s}^{}_{19} \hat{s}^*_{39} & \hat{s}^{}_{27}
	\hat{s}^*_{37} + \hat{s}^{}_{28} \hat{s}^*_{38} + \hat{s}^{}_{29}
	\hat{s}^*_{39} & \frac{1}{2} \left( s^2_{37} +
	s^2_{38} + s^2_{39} \right) \cr \end{matrix} \right) \;,
\nonumber \\
B^{}_2 \hspace{-0.2cm} & \simeq & \hspace{-0.2cm}
I - \left( \begin{matrix} \frac{1}{2} \left( s^2_{17} +
	s^2_{27} + s^2_{37} \right) & 0 & 0 \cr \hat{s}^{*}_{17}
	\hat{s}^{}_{18} + \hat{s}^{*}_{27} \hat{s}^{}_{28} + \hat{s}^{*}_{37}
	\hat{s}^{}_{38} & \frac{1}{2} \left( s^2_{18} + s^2_{28} + s^2_{38}
	\right) & 0 \cr \hat{s}^{*}_{17} \hat{s}^{}_{19} + \hat{s}^{*}_{27}
	\hat{s}^{}_{29} + \hat{s}^{*}_{37} \hat{s}^{}_{39} & \hat{s}^{*}_{18}
	\hat{s}^{}_{19} + \hat{s}^{*}_{28} \hat{s}^{}_{29} + \hat{s}^{*}_{38}
	\hat{s}^{}_{39} & \frac{1}{2} \left( s^2_{19} +
	s^2_{29} + s^2_{39} \right) \cr \end{matrix} \right) \;, \hspace{0.5cm}
\end{eqnarray}
where the terms of ${\cal O}(s^4_{ij})$ (for $i=1,2,3$ and $j=7,8,9$)
have been omitted; and
\begin{eqnarray}
S^{}_2 \simeq -R^\dagger_2 \simeq - \left( \begin{matrix}
\hat{s}^{}_{17} & \hat{s}^{}_{27}
	& \hat{s}^{}_{37} \cr \hat{s}^{}_{18} & \hat{s}^{}_{28} &
	\hat{s}^{}_{38} \cr \hat{s}^{}_{19} & \hat{s}^{}_{29} &
	\hat{s}^{}_{39} \cr \end{matrix} \right) \;,
\end{eqnarray}
where the terms of ${\cal O}(s^3_{ij})$ (for $i=1,2,3$ and $j=7,8,9$) have been omitted.

\end{document}